\documentclass[english,superscriptaddress]{revtex4-1}
\usepackage[T1]{fontenc}
\usepackage[latin9]{inputenc}
\usepackage{color}
\usepackage{babel}
\usepackage{verbatim}
\usepackage{amsmath}
\usepackage{amssymb}
\usepackage{graphicx}
\usepackage[unicode=true,pdfusetitle,
 bookmarks=true,bookmarksnumbered=false,bookmarksopen=false,
 breaklinks=false,pdfborder={0 0 1},backref=false,colorlinks=true]
 {hyperref}

\makeatletter
\allowdisplaybreaks

\@ifundefined{showcaptionsetup}{}{%
 \PassOptionsToPackage{caption=false}{subfig}}
\usepackage{subfig}
\makeatother

\begin{document}
\title{The spin alignment of rho mesons in a pion gas}
\author{Yi-Liang Yin}
\affiliation{Department of Modern Physics, University of Science and Technology
of China, Hefei, Anhui 230026, China}
\author{Wen-Bo Dong}
\affiliation{Department of Modern Physics, University of Science and Technology
of China, Hefei, Anhui 230026, China}
\author{Jin-Yi Pang}
\affiliation{College of Science, University of Shanghai for Science and Technology,
Shanghai 200093, China}
\author{Shi Pu}
\affiliation{Department of Modern Physics, University of Science and Technology
of China, Hefei, Anhui 230026, China}
\author{Qun Wang}
\affiliation{Department of Modern Physics, University of Science and Technology
of China, Hefei, Anhui 230026, China}
\affiliation{School of Mechanics and Physics, Anhui University of Science and Technology,
Huainan,Anhui 232001, China}
\begin{abstract}
We study the spin alignment of neutral rho mesons in a pion gas using
spin kinetic or Boltzmann equations. The $\rho\pi\pi$ coupling is
given by the chiral effective theory. The collision terms at the leading
and next-to-leading order in spin Boltzmann equations are derived.
The evolution of the spin density matrix of the neutral rho meson
is simulated with different initial conditions. The numerical results
show that the interaction of pions and neutral rho mesons creates
very small spin alignment in the central rapidity region if there
is no rho meson in the system at the initial time. Such a small spin
alignment in the central rapidity region will decay rapidly toward
zero in later time. If there are rho mesons with a sizable spin alignment
at the initial time the spin alignment will also decrease rapidly.
We also considered the effect on $\rho_{00}$ from the elliptic flow
of pions in the blast wave model. With vanishing spin alignment at
the initial time, the deviation of $\rho_{00}$ from 1/3 is positive
but very small.
\end{abstract}
\maketitle

\section{Introduction}

The orbital angular momentum and spin are intrinsically connected
with each other, as demonstrated in the Barnett effect \citep{RevModPhys.7.129}
and Einstein-de-Haas effect \citep{dehaas:1915} in materials. In
peripheral collisions of heavy ions, a part of the orbital angular
momentum (OAM) in the initial state can be distributed into the strong
interaction matter via spin-orbit couplings in the form of the hadron's
spin polarization with respect to the direction of OAM (reaction plane),
which is called the global polarization \citep{Liang:2004ph,Liang:2004xn,Betz:2007kg,Gao:2007bc,Becattini:2007sr}.
The spin polarization of hyperons can be measured through their weak
decays in which the parity symmetry is broken \citep{PhysRevLett.36.1113}.
The global polarization of $\Lambda$ hyperons (including anti-paricles)
has been measured by STAR collaboration in Au+Au collisions at 3-200
GeV \citep{STAR:2017ckg,STAR:2018gyt}, by HADES collaboration in
Au+Au and Ag+Ag collisions at 2.42-2.55 GeV \citep{HADES:2022enx}
and by ALICE collaboration in Pb+Pb collisions at 5.02 TeV \citep{ALICE:2021pzu}.
The global polarization of $\Xi$ and $\Omega$ hyperons (including
anti-particles) has also been measured by STAR collaboration in Au+Au
collisions at 200 GeV \citep{STAR:2020xbm}. These experimental measurements
have been explained by various theoretical models (mainly hydrodynamics
and transport models) \citep{Karpenko:2016jyx,Li:2017slc,Xie:2017upb,Sun:2017xhx,Baznat:2017jfj,Shi:2017wpk,Xia:2018tes,Wei:2018zfb,Fu:2020oxj,Ryu:2021lnx,Fu:2021pok,Deng:2021miw,Becattini:2021iol,Wu:2022mkr}.
We refer the readers to some recent review articles in this field
\citep{Wang:2017jpl,Florkowski:2018fap,Gao:2020lxh,Huang:2020dtn,Gao:2020vbh,Becattini:2020ngo,Becattini:2022zvf}.

Most vector mesons decay through strong interaction that preserves
the parity symmetry, so the spin polarization of vector mesons cannot
be measured in the same way as hyperons. The spin density matrix $\rho_{\lambda_{1}\lambda_{2}}$
for the spin-1 vector meson is a $3\times3$ complex matrix with unit
trace, $\mathrm{tr}\rho=1$, where $\lambda_{1}$ and $\lambda_{2}=0,\pm1$
denote the spin states along the spin quantization direction. The
00-element $\rho_{00}$ for the vector meson can be measured by the
angular distribution of its decay product or daugther particle \citep{Schilling:1969um,Liang:2004xn,Yang:2017sdk,Tang:2018qtu},
so $\rho_{00}-1/3$ is an observable that can describe the spin alignment
of the vector meson. If $\rho_{00}=1/3$, the angular distribution
of the daughter particle is isotropic and the vector meson has no
spin alignment. If $\rho_{00}>1/3$, the polarization vector of the
meson is aligned more in the spin quantization direction. If $\rho_{00}<1/3$,
the polarization vector of the meson is aligned more in the transverse
direction perpendicular to the spin quantization direction. The global
spin alignment of $\phi$ and $K^{0*}$ mesons has recently been measured
by STAR collaboration \citep{STAR:2022fan}. It is found that $\rho_{00}^{\phi}$
is significantly larger than 1/3 at lower energies, while $\rho_{00}^{K^{0*}}$
is consistent with 1/3.


There are many sources to the spin alignment of vector mesons \citep{Yang:2017sdk,Xia:2020tyd,Gao:2021rom,Muller:2021hpe,Li:2022vmb,Wagner:2022gza,Kumar:2022ylt,Dong:2023cng,Kumar:2023ghs,Gao:2023wwo}.
In Ref. \citep{Sheng:2019kmk}, some of us proposed that a large deviation
of $\rho_{00}$ from 1/3 for $\phi$ mesons may possibly come from
the $\phi$ field, a strong force field with vacuum quantum number
induced by the current of pseudo-Goldstone bosons. Such a proposal
is based on a nonrelativistic quark coalescence model for the spin
density matrix of vector mesons \citep{Yang:2017sdk,Sheng:2019kmk},
which is only valid for static vector mesons. In Ref. \citep{Sheng:2022ffb},
the relativistic version of the quark coalescence model has been constructed
based on the spin Boltzmann equation with collisions. The model is
successful in describing the experimental data for $\rho_{00}$ for
$\phi$ mesons \citep{Sheng:2022wsy}. Recently some of us made a
prediction for the rapidity dependence of the spin alignment with
the same set of parameters \citep{Sheng:2023urn}, which was later
confirmed by the preliminary data of STAR \citep{Xi:2023quarkmatter}.
We refer the readers to some recent review articles about the spin
alignment of vector mesons \citep{Chen:2023hnb,Wang:2023fvy,Sheng:2023chinphyb}.


In this paper, we try to study the spin alignment of the $\rho^{0}$
meson in a pion gas. As is well-known, the lifetime of the $\rho^{0}$
meson is very short and mainly decays inside the medium. As the result,
the interaction between $\rho^{0}$ and $\pi^{\pm}$ mesons in the
hadron phase of heavy-ion collisions has significant impact on the
spin alignment of the rho meson. This is very different from the $\phi$
meson which is mainly formed by hadronization of quarks. This study
is relevant to the search for the chiral magnetic effect (CME) \citep{Kharzeev:2004ey,Kharzeev:2007jp,Fukushima:2008xe}
since the decay of $\rho^{0}$ to $\pi^{\pm}$ provides a significant
contribution to the background in the $\gamma$ correlator \citep{STAR:2013ksd,STAR:2013zgu,Wang:2016iov}
and the spin alignment of $\rho^{0}$ may have an effect on CME observables
\citep{Tang:2019pbl,Shen:2022gtl}.


The paper is organized as follows. In Sec. \ref{sec:Lagrangain},
an effective Lagrangian is given for the $\rho\pi\pi$ coupling \citep{Fujiwara:1984pk}.
In Sec. \ref{sec:Green's-function-and}, from the Kadanoff-Baym (KB)
equation for Green's functions for pseudoscalar and vector mesons
in the closed-time-path (CTP) formalism \citep{Kadanoff2018QuantumSM},
we derive the spin Boltzmann equations for vector mesons with collisions
\citep{Sheng:2022ffb}. In Sec. \ref{sec:Collision-terms}, we derive
the collision terms at the leading order (LO) and next-to-leading
order (NLO) with the medium effect. The numerical results are given
in Sec. \ref{sec:Numerical-result}. In the final section, Sec. \ref{sec:Conclusions-and-discussions},
are the conclusion and discussion.


The sign convention for the metric tensor is $g_{\mu\nu}=g^{\mu\nu}=\mathrm{diag}\left(1,-1,-1,-1\right)$,
where we use Greek letters to denote four-dimension indices of vectors
or tensors. The four-momentum is defined as $p=p^{\mu}=\left(p^{0},\mathbf{p}\right)$
and $p_{\mu}=\left(p^{0},-\mathbf{p}\right)$, where $p^{0}$ is the
particle's energy. For an on-shell particle, we have $p^{0}=E_{p}=\sqrt{\mathbf{p}^{2}+m^{2}}$.


\section{Effective Lagrangian}

\label{sec:Lagrangain}We consider the chiral effective theory with
SU(2) flavor symmetry. The $\rho$ meson is introduced via the hidden
gauge field. The effective Lagrangian for a system of $\rho^{0}$,
$\pi^{+}$ and $\pi^{-}$ mesons reads
\begin{eqnarray}
\mathcal{L} & = & \mathcal{L}_{\rho}+\mathcal{L}_{\pi}+\mathcal{L}_{\mathrm{int}},\label{eq:Lagrangian}
\end{eqnarray}
where $\mathcal{L}_{\rho}$, $\mathcal{L}_{\pi}$ and $\mathcal{L}_{\mathrm{int}}$
are the Lagrangians for free $\rho^{0}$, free $\pi^{\pm}$, and their
interaction, respectively. They are given by
\begin{align}
\mathcal{L}_{\rho}= & -\frac{1}{4}F_{\mu\nu}F^{\mu\nu}+\frac{1}{2}m_{\rho}^{2}A_{\mu}A^{\mu},\nonumber \\
\mathcal{L}_{\pi}= & \partial_{\mu}\phi^{\dagger}\partial^{\mu}\phi-m_{\pi}^{2}\phi^{\dagger}\phi,\nonumber \\
\mathcal{L}_{\mathrm{int}}= & ig_{\rho\pi\pi}A^{\mu}\Big(\phi^{\dagger}\partial_{\mu}\phi-\phi\partial_{\mu}\phi^{\dagger}\Big),
\end{align}
where $A_{\mu}$ is the real vector field for $\rho^{0}$, $F_{\mu\nu}=\partial_{\mu}A_{\nu}-\partial_{\nu}A_{\mu}$
is the field strength tensor, $m_{\rho}=770$ MeV and $m_{\pi}=139$
MeV are masses of the rho meson and pion respectively, $\phi$ ($\phi^{\dagger}$)
denotes the complex scalar field for $\pi^{+}$ ($\pi^{-}$), and
$g_{\rho\pi\pi}\approx5.9$ is the coupling constant for the $\rho\pi\pi$
vertex. The Lagrangian (\ref{eq:Lagrangian}) is our starting point
to derive the collision terms.


\section{Wigner functions and spin Boltzmann equation}

\label{sec:Green's-function-and}In this section we will introduce
Wigner functions and spin kinetic or Boltzmann equations for vector
mesons. The spin kinetic or Boltzmann equations can be derived from
the KB equation in the CTP formalism \citep{Martin:1959jp,Keldysh:1964ud,Kadanoff2018QuantumSM,Chou:1984es,Blaizot:2001nr,Berges:2004yj,Cassing:2008nn,Cassing:2021fkc}.
The spin kinetic or Boltzmann equations with collision terms are recent
focus and have been derived for spin-1/2 massive fermions \citep{Yang:2020hri,Sheng:2021kfc}
and for vector mesons \citep{Sheng:2022ffb,Sheng:2022wsy,Wagner:2023cct}
in the CTP formalism. They can also be derived in other methods for
spin-1/2 massive fermions \citep{Weickgenannt:2019dks,Li:2019qkf,Sheng:2022ssd,Weickgenannt:2020aaf,Weickgenannt:2021cuo,Lin:2021mvw,Lin:2022tma,Wagner:2022amr}
and for vector mesons \citep{Wagner:2023cct}. The building blocks
of kinetic or Boltzmann equations are Wigner functions in phase space
that are defined from two-point Green's functions \citep{Vasak:1987um,Heinz:1983nx,Blaizot:2001nr,Wang:2001dm,Gao:2012ix,Chen:2012ca,Becattini:2013fla,Gao:2019znl,Weickgenannt:2019dks,Hattori:2019ahi,Wang:2019moi,Weickgenannt:2020aaf,Yang:2020hri,Liu:2020flb,Weickgenannt:2021cuo,Sheng:2021kfc},
see, e.g., Refs. \citep{Gao:2020pfu,Hidaka:2022dmn} for recent reviews.


The real vector and complex scalar fields can be quantized as
\begin{eqnarray}
A^{\mu}(x) & = & \sum_{\lambda=0,\pm1}\int\frac{d^{3}p}{\left(2\pi\hbar\right)^{3}2E_{p}^{\rho}}\nonumber \\
 &  & \times\left[\epsilon^{\mu}(\lambda,{\bf p})a_{V}(\lambda,{\bf p})e^{-ip\cdot x/\hbar}+\epsilon^{\mu\ast}(\lambda,{\bf p})a_{V}^{\dagger}(\lambda,{\bf p})e^{ip\cdot x/\hbar}\right],\label{eq:vector field}\\
\phi(x) & = & \int\frac{d^{3}k}{\left(2\pi\hbar\right)^{3}2E_{k}^{\pi}}\left[a({\bf k})e^{-ik\cdot x/\hbar}+b^{\dagger}({\bf k})e^{ik\cdot x/\hbar}\right],\label{eq:scalar field}
\end{eqnarray}
where $E_{p}^{\rho}=\sqrt{\mathbf{p}^{2}+m_{\rho}^{2}}$ and $E_{k}^{\pi}=\sqrt{\mathbf{k}^{2}+m_{\pi}^{2}}$
are the energies of $\rho$ and $\pi$ respectively, $\lambda$ denotes
the spin state with respect to the spin quantization direction, and
$\epsilon^{\mu}(\lambda,{\bf p})$ is the polarization vector
\begin{eqnarray}
\epsilon^{\mu}(\lambda,\mathbf{p}) & = & \left(\frac{\mathbf{p}\cdot\boldsymbol{\epsilon}_{\lambda}}{m_{\rho}},\boldsymbol{\epsilon}_{\lambda}+\frac{\mathbf{p}\cdot\boldsymbol{\epsilon}_{\lambda}}{m_{\rho}(E_{p}+m_{\rho})}\mathbf{p}\right),
\end{eqnarray}
with $\boldsymbol{\epsilon}_{\lambda}$ being the polarization three-vector
of the vector meson in its rest frame and given by
\begin{align}
\boldsymbol{\epsilon}_{0}= & (0,1,0),\nonumber \\
\boldsymbol{\epsilon}_{+1}= & -\frac{1}{\sqrt{2}}(i,0,1),\nonumber \\
\boldsymbol{\epsilon}_{-1}= & \frac{1}{\sqrt{2}}(-i,0,1).
\end{align}
Here $\boldsymbol{\epsilon}_{0}$ is the spin quantization direction
and is chosen to be $+y$ direction. The polarization vector $\epsilon^{\mu}(\lambda,{\bf p})$
has following properties
\begin{align}
 & p_{\mu}\epsilon^{\mu}(\lambda,{\bf p})=0\nonumber \\
 & \epsilon(\lambda,{\bf p})\cdot\epsilon^{*}(\lambda^{\prime},{\bf p})=-\delta_{\lambda\lambda^{\prime}}\nonumber \\
 & \sum_{\lambda}\epsilon^{\mu}(\lambda,{\bf p})\epsilon^{\nu*}(\lambda,{\bf p})=-\left(g^{\mu\nu}-\frac{p^{\mu}p^{\nu}}{m_{\rho}^{2}}\right).\label{eq:polar-vector-rel}
\end{align}


Then we can define the two-point Green's functions on the CTP for
the vector and pseudoscalar meson,
\begin{eqnarray}
G_{CTP}^{\mu\nu}(x_{1},x_{2}) & = & \left\langle T_{C}A^{\mu}(x_{1})A^{\nu}(x_{2})\right\rangle ,\label{eq:G_vector}\\
S_{CTP}(x_{1},x_{2}) & = & \left\langle T_{C}\phi(x_{1})\phi^{\dagger}(x_{2})\right\rangle .\label{eq:G_scalar}
\end{eqnarray}
The two-point Green's functions $G_{\mu\nu}^{\lessgtr}$ for the vector
meson at the leading order are given as \citep{Sheng:2022ffb},
\begin{eqnarray}
G_{\mu\nu}^{<}(x,p) & = & 2\pi\hbar\sum_{\lambda_{1},\lambda_{2}}\delta\left(p^{2}-m_{\rho}^{2}\right)\left\{ \theta(p^{0})\epsilon_{\mu}\left(\lambda_{1},{\bf p}\right)\epsilon_{\nu}^{\ast}\left(\lambda_{2},{\bf p}\right)f_{\lambda_{1}\lambda_{2}}(x,{\bf p})\right.\nonumber \\
 &  & \left.+\theta(-p^{0})\epsilon_{\mu}^{\ast}\left(\lambda_{1},-{\bf p}\right)\epsilon_{\nu}\left(\lambda_{2},-{\bf p}\right)\left[\delta_{\lambda_{2}\lambda_{1}}+f_{\lambda_{2}\lambda_{1}}(x,-{\bf p})\right]\right\} ,\label{eq:G_less}\\
G_{\mu\nu}^{>}(x,p) & = & 2\pi\hbar\sum_{\lambda_{1},\lambda_{2}}\delta\left(p^{2}-m_{\rho}^{2}\right)\left\{ \theta(p^{0})\epsilon_{\mu}\left(\lambda_{1},{\bf p}\right)\epsilon_{\nu}^{\ast}\left(\lambda_{2},{\bf p}\right)\left[\delta_{\lambda_{1}\lambda_{2}}+f_{\lambda_{1}\lambda_{2}}(x,{\bf p})\right]\right.\nonumber \\
 &  & \left.+\theta(-p^{0})\epsilon_{\mu}^{\ast}\left(\lambda_{1},-{\bf p}\right)\epsilon_{\nu}\left(\lambda_{2},-{\bf p}\right)f_{\lambda_{2}\lambda_{1}}(x,-{\bf p})\right\} ,\label{eq:G_larger}
\end{eqnarray}
where $f_{\lambda_{1}\lambda_{2}}(x,{\bf p})$ is the matrix valued
spin dependent distribution (MVSD) for the rho meson,
\begin{equation}
f_{\lambda_{1}\lambda_{2}}(x,{\bf p})\equiv\int\frac{d^{4}u}{2(2\pi\hbar)^{3}}\delta(p\cdot u)e^{-iu\cdot x/\hbar}\left\langle a_{\rho}^{\dagger}\left(\lambda_{2},{\bf p}-\frac{{\bf u}}{2}\right)a_{\rho}\left(\lambda_{1},{\bf p}+\frac{{\bf u}}{2}\right)\right\rangle .
\end{equation}
One can check that $f_{\lambda_{1}\lambda_{2}}(x,{\bf p})$ is an
Hermitian matrix, $f_{\lambda_{1}\lambda_{2}}^{*}(x,{\bf p})=f_{\lambda_{2}\lambda_{1}}(x,{\bf p})$.
The two-point Green's function for $\pi^{\pm}$ at the leading order
is
\begin{align}
S^{<}(x,k)= & 2\pi\hbar\delta\left(k^{2}-m_{\pi}^{2}\right)\nonumber \\
 & \times\left\{ \theta(k^{0})f_{\pi^{+}}(x,\mathbf{k})+\theta(-k^{0})\left[1+f_{\pi^{-}}(x,-\mathbf{k})\right]\right\} ,\\
S^{>}(x,k)= & 2\pi\hbar\delta\left(k^{2}-m_{\pi}^{2}\right)\nonumber \\
 & \times\left\{ \theta(k^{0})\left[1+f_{\pi^{+}}(x,\mathbf{k})\right]+\theta(-k^{0})f_{\pi^{-}}(x,-\mathbf{k})\right\} ,
\end{align}
where $f_{\pi^{\pm}}(x,\mathbf{p})$ is the distribution for $\pi^{\pm}$.
For notational convenience, we use $G$ and $p$ to denote the Green's
function and momentum for the rho meson respectively, while we use
$S$ and $k$ to denote the Green's function and momentum for $\pi^{\pm}$
respectively.


We start from the KB equation to derive the spin Boltzmann equation
for the vector meson \citep{Sheng:2022ffb}
\begin{eqnarray}
 &  & p\cdot\partial_{x}G^{<,\mu\nu}(x,p)-\frac{1}{4}\left[p^{\mu}\partial_{\eta}^{x}G^{<,\eta\nu}(x,p)+p^{\nu}\partial_{\eta}^{x}G^{<,\mu\eta}(x,p)\right]\nonumber \\
 & = & \frac{1}{4}\left[\Sigma_{\;\;\;\;\alpha}^{<,\mu}\left(x,p\right)G^{>,\alpha\nu}\left(x,p\right)-\Sigma_{\;\;\;\;\alpha}^{>,\mu}\left(x,p\right)G^{<,\alpha\nu}\left(x,p\right)\right]\nonumber \\
 &  & +\frac{1}{4}\left[G_{\ \ \ \ \alpha}^{>,\mu}\left(x,p\right)\Sigma^{<,\alpha\nu}\left(x,p\right)-G_{\ \ \ \ \alpha}^{<,\mu}\left(x,p\right)\Sigma^{>,\alpha\nu}\left(x,p\right)\right].\label{eq:boltzmann-eq}
\end{eqnarray}
In the above equation, the Poisson bracket terms are not considered.
Multiplying $\epsilon_{\mu}^{*}\left(\lambda_{1},{\bf p}\right)\epsilon_{\nu}\left(\lambda_{2},{\bf p}\right)$
to both side of Eq. (\ref{eq:boltzmann-eq}) and choose $p_{0}>0$
part, we obtain
\begin{eqnarray}
p\cdot\partial_{x}f_{\lambda_{1}\lambda_{2}}(x,\mathbf{p}) & = & -\frac{1}{4}\delta_{\lambda_{2}\lambda_{2}^{\prime}}\epsilon_{\mu}^{*}\left(\lambda_{1},{\bf p}\right)\epsilon^{\alpha}\left(\lambda_{1}^{\prime},{\bf p}\right)\nonumber \\
 &  & \times\left\{ \left[\delta_{\lambda_{1}^{\prime}\lambda_{2}^{\prime}}+f_{\lambda_{1}^{\prime}\lambda_{2}^{\prime}}(x,{\bf p})\right]\Sigma_{\;\;\;\;\alpha}^{<,\mu}\left(x,p\right)-f_{\lambda_{1}^{\prime}\lambda_{2}^{\prime}}(x,{\bf p})\Sigma_{\;\;\;\;\alpha}^{>,\mu}\left(x,p\right)\right\} \nonumber \\
 &  & -\frac{1}{4}\delta_{\lambda_{1}\lambda_{1}^{\prime}}\epsilon_{\nu}\left(\lambda_{2},{\bf p}\right)\epsilon_{\alpha}^{\ast}\left(\lambda_{2}^{\prime},{\bf p}\right)\nonumber \\
 &  & \times\left\{ \left[\delta_{\lambda_{1}^{\prime}\lambda_{2}^{\prime}}+f_{\lambda_{1}^{\prime}\lambda_{2}^{\prime}}(x,{\bf p})\right]\Sigma^{<,\alpha\nu}\left(x,p\right)-f_{\lambda_{1}^{\prime}\lambda_{2}^{\prime}}(x,{\bf p})\Sigma^{>,\alpha\nu}\left(x,p\right)\right\} .\label{eq:spin-Botzmann-equation}
\end{eqnarray}
The above equation is the spin Boltzmann equation for the vector meson
in terms of MVSDs. The MVSDs of spin-1/2 fermions are defined in Refs.
\citep{Becattini:2013fla,Sheng:2021kfc} and those for vector mesons
are defined in Refs. \citep{Sheng:2022wsy,Sheng:2022ffb}. The spin
density matrix is just the normalized MVSD
\begin{equation}
\rho_{\lambda_{1}\lambda_{2}}=\frac{f_{\lambda_{1}\lambda_{2}}}{\sum_{\lambda}f_{\lambda\lambda}}=\frac{f_{\lambda_{1}\lambda_{2}}}{\mathrm{Tr}f}.
\end{equation}
The spin alignment is given by the 00 element $\rho_{00}$.

We make a few remarks about the spin kinetic or Boltzmann equation
(\ref{eq:spin-Botzmann-equation}). The collision terms in the right-hand
side of Eq. (\ref{eq:spin-Botzmann-equation}) are the result of the
on-shell approximation. In such an approximation, the retarded and
advanced components of self-energies and two-point Green's functions
are neglected so that the collision terms only depend on the ``<''
and ``>'' components. Hence the contributions to the spin density
matrix of vector mesons come from collisions of on-shell particles
including the vector meson's annihilation and production processes.
The contribution from different retarded and advanced self-energies
for transverse and longitudinal modes in equilbrium is called the
off-shell contribution \citep{Kim:2019ybi,Li:2022vmb,Dong:2023cng,Seck:2023oyt},
which belongs to a different kind of the contribution from the one
we consider in this paper.

In the next section we will derive the self-energy $\Sigma_{\mu\nu}$
and then collision terms incorporating the interaction part of the
Lagrangian.

\section{Collision terms}

\label{sec:Collision-terms}For clarification, we decompose the collision
terms, the right-hand-side (r.h.s.) of Eq. (\ref{eq:spin-Botzmann-equation}),
into $C_{\mathrm{coal}/\mathrm{diss}}$ and $C_{\mathrm{scat}}$ for
the coalescence-dissociation and scattering processes respectively,
where $C_{\mathrm{coal}/\mathrm{diss}}$ have contrbutions at LO and
NLO, $C_{\mathrm{coal}/\mathrm{diss}}=C_{\mathrm{coal}/\mathrm{diss}}^{(0)}+C_{\mathrm{coal}/\mathrm{diss}}^{(1)}$,
while $C_{\mathrm{scat}}$ is of NLO. Note that we only consider contributions
up to NLO in this paper. Then Eq. (\ref{eq:spin-Botzmann-equation})
can be written as
\begin{eqnarray}
\frac{p}{E_{p}^{\rho}}\cdot\partial_{x}f_{\lambda_{1}\lambda_{2}}(x,\mathbf{p}) & = & C_{\mathrm{coal}/\mathrm{diss}}+C_{\mathrm{scat}},\label{eq:MVSD-Boltzmann-equation}
\end{eqnarray}
where the spin indices $\lambda_{1}$, $\lambda_{2}$ and phase space
variables $x,\mathbf{p}$ have been suppressed in collision terms.
In this work, for simplicity, we adopt the gradient expansion in space
and neglect spatial gradients of $f_{\lambda_{1}\lambda_{2}}$ at
the leading order. This corresponds to the assumption that the system
is homogeneous in space. So Eq. (\ref{eq:MVSD-Boltzmann-equation})
becomes
\begin{eqnarray}
\partial_{t}f_{\lambda_{1}\lambda_{2}}(x,\mathbf{p}) & = & C_{\mathrm{coal}/\mathrm{diss}}+C_{\mathrm{scat}}.\label{eq:boltzmann-compact}
\end{eqnarray}
We will evaluate $C_{\mathrm{coal}/\mathrm{diss}}$ and $C_{\mathrm{scat}}$
one by one.


\subsection{Leading order}

\begin{figure}
\begin{centering}
\includegraphics[scale=0.5]{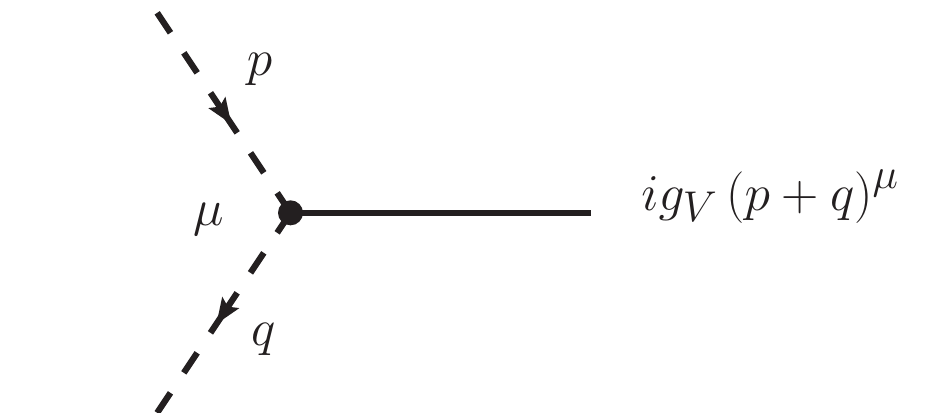}
\par\end{centering}
\caption{The Feynman rule for the $\rho\pi\pi$ vertex, where the solid line
represents $\rho^{0}$'s on-shell state and dashed lines represent
$\pi^{\pm}$'s on-shell states. \label{fig:Feynman-rules}}
\end{figure}

The Feynman rule for the $\rho\pi\pi$ vertex is in Fig. (\ref{fig:Feynman-rules}).
In Feynman diagrams, solid lines represent $\rho^{0}$ meson's on-shell
states (external lines) or propagators (internal lines) and dashed
lines represent $\pi^{\pm}$ meson's on-shell states (external lines)
or propagators (internal lines). The arrow on the $\rho^{0}$ meson's
propagator only labels the momentum direction, since $\rho^{0}$ is
the charge neutral particle, while the arrow on $\pi^{\pm}$ meson's
propagator labels the momentum direction of $\pi^{+}$ or the inverse
momentum direction of $\pi^{-}$.

\begin{figure}
\subfloat[]{\includegraphics[scale=0.45]{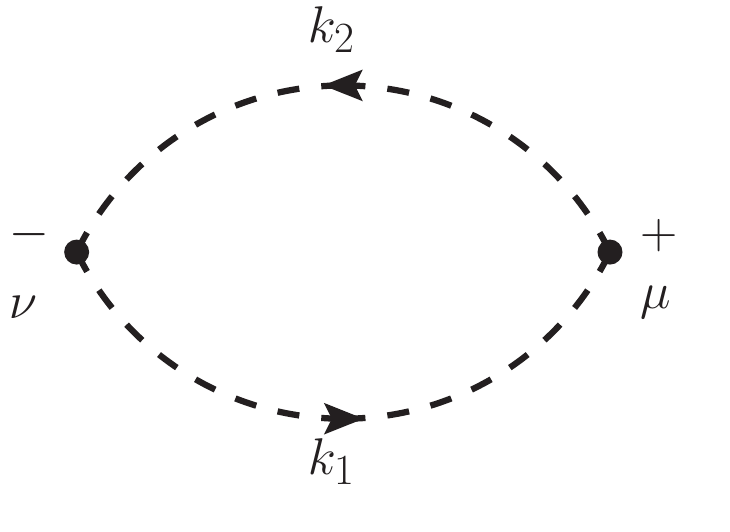}

}\hspace{1cm}\subfloat[]{\includegraphics[scale=0.45]{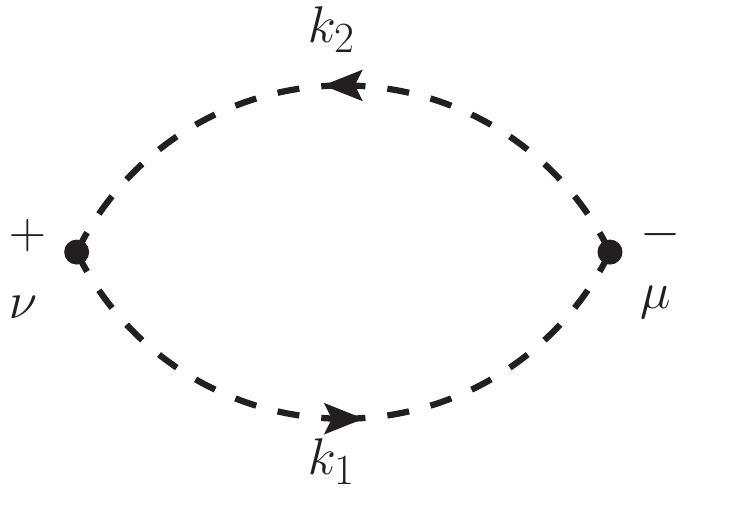}

}

\caption{Leading-order Feynman diagrams for (a) $\Sigma_{\mu\nu}^{<}(x,p)$
and (b) $\Sigma_{\mu\nu}^{>}(x,p)$, where dashed lines represent
propagators of $\pi^{\pm}$ mesons. The external moment $p$ is flowing
from left to right. \label{fig:Leading-order-Feynman-diagrams}}
\end{figure}

The self-energies corresponding to leading order (LO) Feynman diagrams
in Fig. (\ref{fig:Leading-order-Feynman-diagrams}) are given as
\begin{eqnarray}
\Sigma_{\mu\nu}^{<}(x,p) & = & -g_{V}^{2}\int\frac{d^{4}k_{1}}{(2\pi\hbar)^{4}}\int\frac{d^{4}k_{2}}{(2\pi\hbar)^{4}}(2\pi\hbar)^{4}\delta^{(4)}\left(p-k_{1}+k_{2}\right)\nonumber \\
 &  & \times\left(k_{1\mu}+k_{2\mu}\right)\left(k_{1\nu}+k_{2\nu}\right)S^{<}(x,k_{1})S^{>}(x,k_{2}),\label{eq:Sigma_less_1}\\
\Sigma_{\mu\nu}^{>}(x,p) & = & -g_{V}^{2}\int\frac{d^{4}k_{1}}{(2\pi\hbar)^{4}}\int\frac{d^{4}k_{2}}{(2\pi\hbar)^{4}}\left(2\pi\hbar\right)^{4}\delta^{(4)}\left(p-k_{1}+k_{2}\right)\nonumber \\
 &  & \times\left(k_{1\mu}+k_{2\mu}\right)\left(k_{1\nu}+k_{2\nu}\right)S^{>}(x,k_{1})S^{<}(x,k_{2}).\label{eq:Sigma_larger_1}
\end{eqnarray}

In deriving Eq. (\ref{eq:spin-Botzmann-equation}), we have chosen
$p^{0}>0$, so $k_{1}^{0}$ and $k_{2}^{0}$ must satisfy $k_{1}^{0}>0$
and $k_{2}^{0}<0$, which means the on-shell process $\rho^{0}\leftrightarrow\pi^{+}\pi^{-}$
is allowed but $\pi^{\pm}\leftrightarrow\rho^{0}\pi^{\pm}$ is forbidden.
The discussion about the sign of $k_{1}^{0}$ and $k_{2}^{0}$ can
be found in Ref. \citep{Sheng:2022ffb}.


Consequently, the LO self-energies in (\ref{eq:Sigma_less_1}) and
(\ref{eq:Sigma_larger_1}) can be put into the form
\begin{eqnarray}
\Sigma_{\mu\nu}^{<}(x,p) & = & -g_{V}^{2}\int\frac{d^{3}k_{1}}{(2\pi\hbar)^{3}2E_{k_{1}}^{\pi}}\int\frac{d^{3}k_{2}}{(2\pi\hbar)^{3}2E_{k_{2}}^{\pi}}(2\pi\hbar)^{4}\delta^{(4)}\left(p-k_{1}-k_{2}\right)\nonumber \\
 &  & \times\left(k_{1\mu}-k_{2\mu}\right)\left(k_{1\nu}-k_{2\nu}\right)f_{\pi^{+}}(x,\mathbf{k}_{1})f_{\pi^{-}}(x,\mathbf{k}_{2}),\label{eq:Sigma_less_2}\\
\Sigma_{\mu\nu}^{>}(x,p) & = & -g_{V}^{2}\int\frac{d^{3}k_{1}}{(2\pi\hbar)^{3}2E_{k_{1}}^{\pi}}\int\frac{d^{3}k_{2}}{(2\pi\hbar)^{3}2E_{k_{2}}^{\pi}}(2\pi\hbar)^{4}\delta^{(4)}\left(p-k_{1}-k_{2}\right)\nonumber \\
 &  & \times\left(k_{1\mu}-k_{2\mu}\right)\left(k_{1\nu}-k_{2\nu}\right)\left[1+f_{\pi^{+}}(x,\mathbf{k}_{1})\right]\left[1+f_{\pi^{-}}(x,\mathbf{k}_{2})\right].\label{eq:Sigma_larger_2}
\end{eqnarray}
Substituting above equations into Eq. (\ref{eq:spin-Botzmann-equation}),
we obtain
\begin{eqnarray}
C_{\mathrm{coal}/\mathrm{diss}}^{(0)}\left(\rho^{0}\leftrightarrow\pi^{+}\pi^{-}\right) & = & \frac{g_{V}^{2}}{E_{p}^{\rho}}\int\frac{d^{3}k}{(2\pi\hbar)^{3}4E_{k}^{\pi}E_{p-k}^{\pi}}2\pi\hbar\delta\left(E_{p}^{\rho}-E_{k}^{\pi}-E_{p-k}^{\pi}\right)\nonumber \\
 &  & \times\left[\delta_{\lambda_{2}\lambda_{2}^{\prime}}k\cdot\epsilon^{*}(\lambda_{1},\mathbf{p})k\cdot\epsilon(\lambda_{1}^{\prime},\mathbf{p})+\delta_{\lambda_{1}\lambda_{1}^{\prime}}k\cdot\epsilon(\lambda_{2},\mathbf{p})k\cdot\epsilon^{*}(\lambda_{2}^{\prime},\mathbf{p})\right]\nonumber \\
 &  & \times\left\{ f_{\pi^{+}}(x,\mathbf{k})f_{\pi^{-}}(x,{\bf p}-\mathbf{k})\left[\delta_{\lambda_{1}^{\prime}\lambda_{2}^{\prime}}+f_{\lambda_{1}^{\prime}\lambda_{2}^{\prime}}(x,{\bf p})\right]\right.\nonumber \\
 &  & \left.-\left[1+f_{\pi^{+}}(x,\mathbf{k})\right]\left[1+f_{\pi^{-}}(x,{\bf p}-\mathbf{k})\right]f_{\lambda_{1}^{\prime}\lambda_{2}^{\prime}}(x,{\bf p})\right\} ,
\end{eqnarray}
where we have used Eq. (\ref{eq:polar-vector-rel}).


\subsection{Next-to-leading order}

\begin{figure}
\subfloat[]{\includegraphics[scale=0.5]{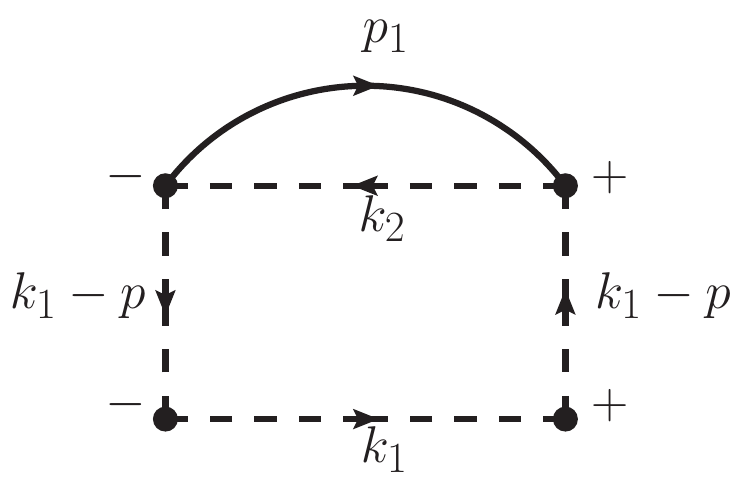}

}\subfloat[]{\includegraphics[scale=0.5]{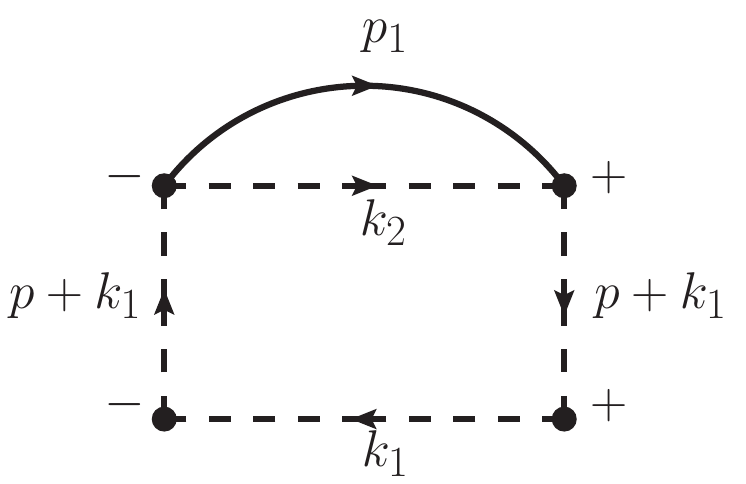}

}

\subfloat[]{\includegraphics[scale=0.5]{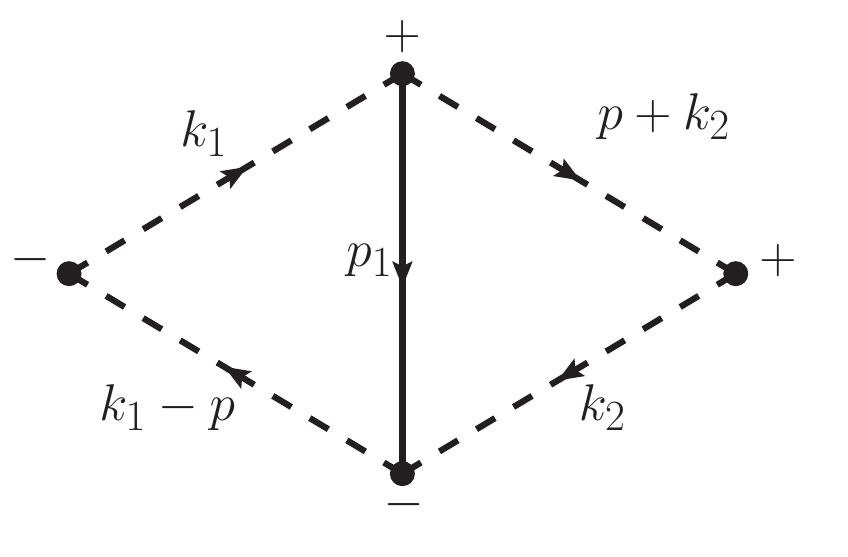}

}\subfloat[]{\includegraphics[scale=0.5]{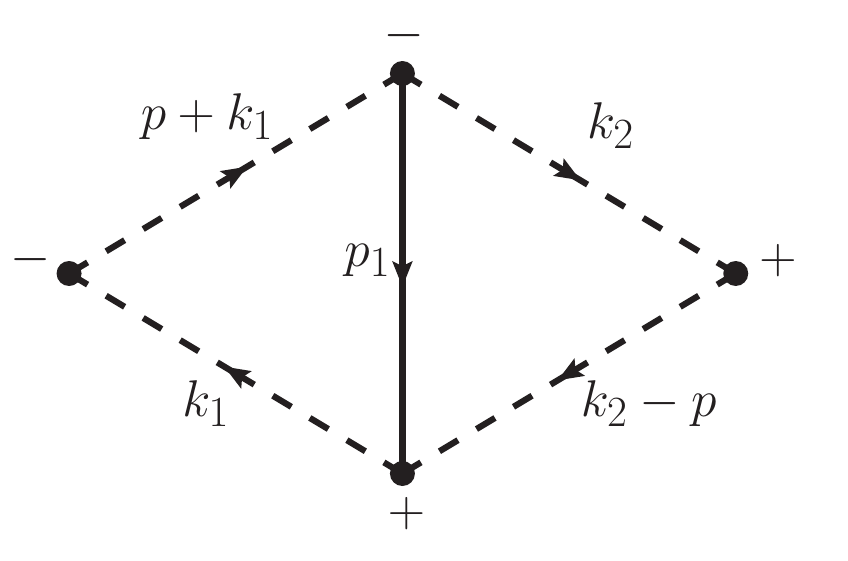}

}

\caption{Feynman diagrams for $\Sigma_{\mu\nu}^{<}(x,p)$ at the next-to-leading
order. The solid lines represent $\rho^{0}$ meson's propagators and
dashed lines represent the propagators of $\pi^{\pm}$ mesons. The
external momentum $p$ is flowing from left to right. \label{fig:Next-to-leading-order-Feynman}}
\end{figure}

The Feynman diagrams for $\Sigma^{<}(x,p)$ at next-to-leading order
(NLO) are shown in Fig. (\ref{fig:Next-to-leading-order-Feynman}).
Considering the difference between $\Sigma^{<}(x,p)$ and $\Sigma^{>}(x,p)$
is to interchange between the positive and negative branch, we can
evaluate $\Sigma^{<}(x,p)$ first and then replace $\lessgtr$ with
$\gtrless$ in $\Sigma^{<}(x,p)$ to obtain $\Sigma^{>}(x,p)$. The
free pion's Feynman propagators with time and reverse-time order are
\begin{eqnarray}
S^{F}(k) & = & \frac{i}{k^{2}-m_{\pi}^{2}},\\
S^{\overline{F}}(k) & = & \frac{-i}{k^{2}-m_{\pi}^{2}}.
\end{eqnarray}
The medium corrections for $S^{F}$ and $S^{\overline{F}}$ will be
discussed in the next subsection.

\begin{figure}
\subfloat[]{\includegraphics[scale=0.6]{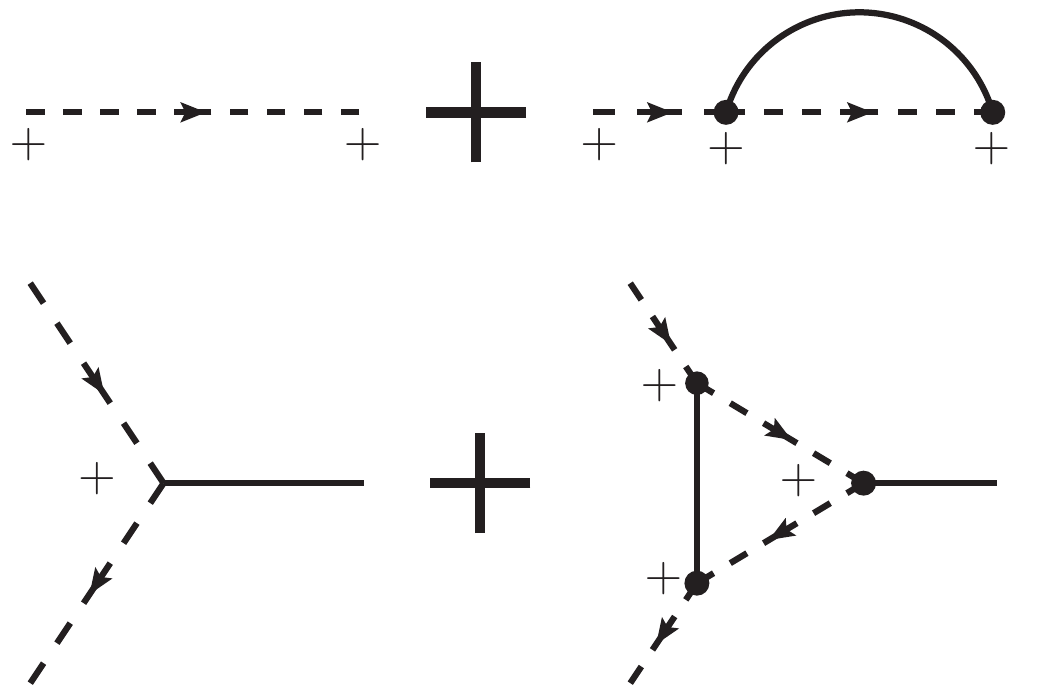}

}

\subfloat[]{\includegraphics[scale=0.6]{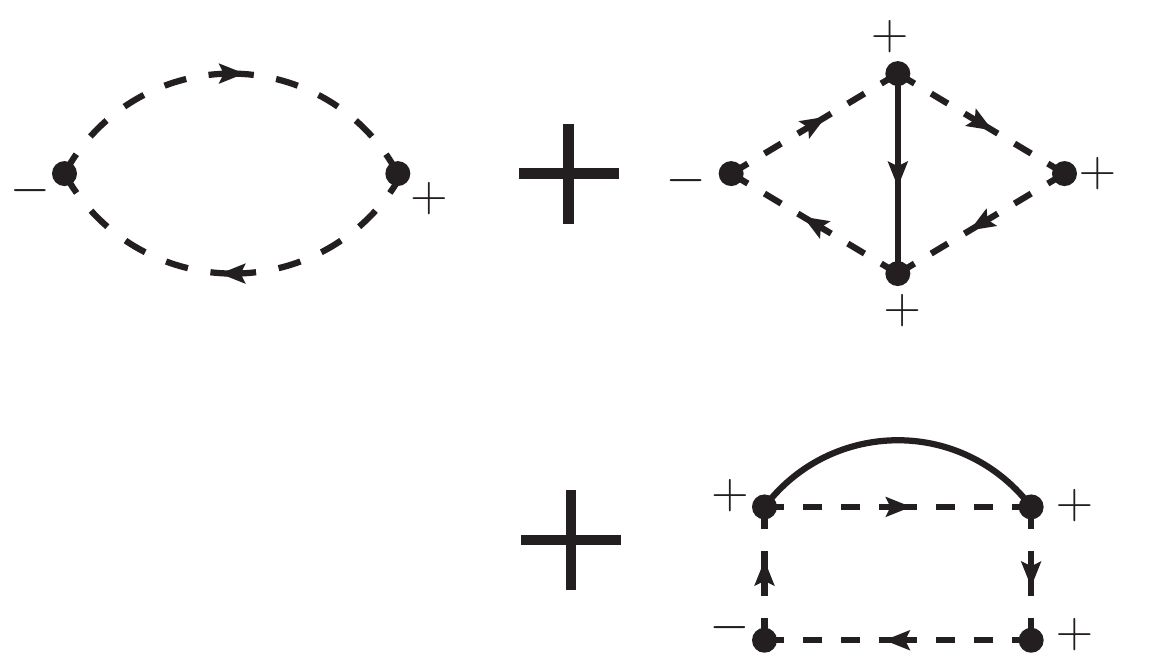}

}

\caption{Examples of propagator and vertex corrections. \label{fig:renormalization-diagrams}}
\end{figure}

We can see that Fig. (\ref{fig:Next-to-leading-order-Feynman})(a)
and (b) are different in orientations of pion loops, and Fig. (\ref{fig:Next-to-leading-order-Feynman})(c)
and (d) are different in time branches for two middle points with
momentum $p_{1}$. In Fig. (\ref{fig:Next-to-leading-order-Feynman})
we choose a particular direction for $p_{1}$ in the vector meson's
propagator, actually one is free to choose any direction without changing
the final result. Other combinations of time branches for upper vertices
in Fig. (\ref{fig:Next-to-leading-order-Feynman})(a) and (b) and
middle vertices in Fig. (\ref{fig:Next-to-leading-order-Feynman})(c)
and (d) correspond to loop corrections to propagators and vertices
respectively, which need renormalization as in quantum field theory
in vacuum. For example, in Fig. (\ref{fig:Next-to-leading-order-Feynman})(a),
other combinations of time branches for two upper vertices (from left
to right) are $++$ and $--$, which correspond to the loop correction
to the right and left pion propagator respectively, as shown in Fig.
\ref{fig:renormalization-diagrams}. As another example, in Fig. (\ref{fig:Next-to-leading-order-Feynman})(c),
other combinations of time branches for two upper vertices (from left
to right) are $++$ and $--$, which correspond to the loop correction
to the right and left vertex respectively, as shown in Fig. \ref{fig:renormalization-diagrams}.

Now we can obtain collision terms at NLO. The result has three parts
corresponding to three processes, $\rho^{0}\pi^{+}\leftrightarrow\rho^{0}\pi^{+}$,
$\rho^{0}\pi^{-}\leftrightarrow\rho^{0}\pi^{-}$ and $\rho^{0}\rho^{0}\leftrightarrow\pi^{+}\pi^{-}$,
\begin{eqnarray}
C_{\mathrm{scat}}\left(\rho^{0}\pi^{\pm}\leftrightarrow\rho^{0}\pi^{\pm}\right) & = & \frac{4g_{V}^{4}}{E_{p}^{\rho}}\int\frac{d^{3}k_{1}}{\left(2\pi\hbar\right)^{3}2E_{k_{1}}^{\pi}}\int\frac{d^{3}k_{2}}{\left(2\pi\hbar\right)^{3}2E_{k_{2}}^{\pi}}\int\frac{d^{3}p_{1}}{\left(2\pi\hbar\right)^{3}2E_{p_{1}}^{\rho}}\nonumber \\
 &  & \times\left(2\pi\hbar\right)^{4}\delta^{(4)}\left(p+k_{2}-p_{1}-k_{1}\right)\nonumber \\
 &  & \times\left[\delta_{\lambda_{2}\lambda_{2}^{\prime}}D_{(1)}(s_{1},\lambda_{1})D_{(1)}^{*}(s_{2},\lambda_{1}^{\prime})+\delta_{\lambda_{1}\lambda_{1}^{\prime}}D_{(1)}(s_{1},\lambda_{2}^{\prime})D_{(1)}^{*}(s_{2},\lambda_{2})\right]\nonumber \\
 &  & \times\left[f_{s_{1}s_{2}}(x,\mathbf{p}_{1})f_{\pi^{\pm}}(x,\mathbf{k}_{1})\left(1+f_{\pi^{\pm}}(x,\mathbf{k}_{2})\right)\left(\delta_{\lambda_{1}^{\prime}\lambda_{2}^{\prime}}+f_{\lambda_{1}^{\prime}\lambda_{2}^{\prime}}(x,{\bf p})\right)\right.\nonumber \\
 &  & \left.-\left(\delta_{s_{1}s_{2}}+f_{s_{1}s_{2}}(x,\mathbf{p}_{1})\right)\left(1+f_{\pi^{\pm}}(x,\mathbf{k}_{1})\right)f_{\pi^{\pm}}(x,\mathbf{k}_{2})f_{\lambda_{1}^{\prime}\lambda_{2}^{\prime}}(x,{\bf p})\right],\label{eq:rho-pi-scattering_1}
\end{eqnarray}
\begin{eqnarray}
C_{\mathrm{coal}/\mathrm{diss}}^{(1)}\left(\rho^{0}\rho^{0}\leftrightarrow\pi^{+}\pi^{-}\right) & = & \frac{4g_{V}^{4}}{E_{p}^{\rho}}\int\frac{d^{3}k_{1}}{\left(2\pi\hbar\right)^{3}2E_{k_{1}}^{\pi}}\int\frac{d^{3}k_{2}}{\left(2\pi\hbar\right)^{3}2E_{k_{2}}^{\pi}}\int\frac{d^{3}p_{1}}{\left(2\pi\hbar\right)^{3}2E_{p_{1}}^{\rho}}\nonumber \\
 &  & \times\left(2\pi\hbar\right)^{4}\delta^{(4)}\left(p+p_{1}-k_{1}-k_{2}\right)\nonumber \\
 &  & \times\left[\delta_{\lambda_{2}\lambda_{2}^{\prime}}D_{(2)}(s_{1},\lambda_{1}^{\prime})D_{(2)}^{*}(s_{2},\lambda_{1})+\delta_{\lambda_{1}\lambda_{1}^{\prime}}D_{(2)}(s_{1},\lambda_{2})D_{(2)}^{*}(s_{2},\lambda_{2}^{\prime})\right]\nonumber \\
 &  & \times\left[f_{\pi^{+}}(x,\mathbf{k}_{1})f_{\pi^{-}}(x,\mathbf{k}_{2})\left(\delta_{s_{1}s_{2}}+f_{s_{1}s_{2}}(x,\mathbf{p}_{1})\right)\left(\delta_{\lambda_{1}^{\prime}\lambda_{2}^{\prime}}+f_{\lambda_{1}^{\prime}\lambda_{2}^{\prime}}(x,{\bf p})\right)\right.\nonumber \\
 &  & \left.-\left(1+f_{\pi^{+}}(x,\mathbf{k}_{1})\right)\left(1+f_{\pi^{-}}(x,\mathbf{k}_{2})\right)f_{s_{1}s_{2}}(x,\mathbf{p}_{1})f_{\lambda_{1}^{\prime}\lambda_{2}^{\prime}}(x,{\bf p})\right],\label{eq:rho-pi-scattering_2}
\end{eqnarray}
where we have used $s_{1}$ and $s_{2}$ to label spin states in propagators
of $\rho^{0}$, used Eq. (\ref{eq:polar-vector-rel}) and the on-shell
condition, and defined
\begin{align}
D_{(1)}(s,\lambda)= & \frac{\left[k_{1}\cdot\epsilon\left(s,\mathbf{p}_{1}\right)\right]\left[k_{2}\cdot\epsilon^{*}\left(\lambda,{\bf p}\right)\right]}{\left(p+k_{2}\right)^{2}-m_{\pi}^{2}}+\frac{\left[k_{2}\cdot\epsilon\left(s,\mathbf{p}_{1}\right)\right]\left[k_{1}\cdot\epsilon^{*}\left(\lambda,{\bf p}\right)\right]}{\left(p-k_{1}\right)^{2}-m_{\pi}^{2}},\nonumber \\
D_{(2)}(s,\lambda)= & \frac{\left[k_{1}\cdot\epsilon\left(s,\mathbf{p}_{1}\right)\right]\left[k_{2}\cdot\epsilon\left(\lambda,{\bf p}\right)\right]}{\left(p-k_{2}\right)^{2}-m_{\pi}^{2}}+\frac{\left[k_{2}\cdot\epsilon\left(s,\mathbf{p}_{1}\right)\right]\left[k_{1}\cdot\epsilon\left(\lambda,{\bf p}\right)\right]}{\left(p-k_{1}\right)^{2}-m_{\pi}^{2}}.\label{eq:pi-propagator}
\end{align}
One can check that the collision terms are Hermitian in consistence
with $f_{\lambda_{1}\lambda_{2}}$.


So far we have completed the derivation of the spin Boltzmann equation
with collision terms at LO and NLO.


\subsection{Regulation of pion propagators}

In the collision term $C_{\mathrm{scat}}\left(\rho^{0}\pi^{\pm}\leftrightarrow\rho^{0}\pi^{\pm}\right)$,
there are pion propagators which may diverge at the pion mass pole.
To regulate these pion propagators, we introduce self-energy corrections
with medium effects as
\begin{eqnarray}
S^{F}(k) & = & \frac{i}{k^{2}-m_{\pi}^{2}-\Sigma^{F}(k)}\label{eq:Feynman-propogator-F}\\
S^{\overline{F}}(k) & = & \frac{-i}{k^{2}-m_{\pi}^{2}+\Sigma^{\overline{F}}(k)},\label{eq:Feynman-propogator-F_bar}
\end{eqnarray}
where $\Sigma^{F}$ is the self-energy for pions. The real part of
the self-energy gives the mass correction, while the imaginary part
is associated with the medium effect. In this work, we only consider
the imaginary part of the self-energy since the mass correction from
the real part is much smaller.

\begin{figure}
\includegraphics[scale=0.5]{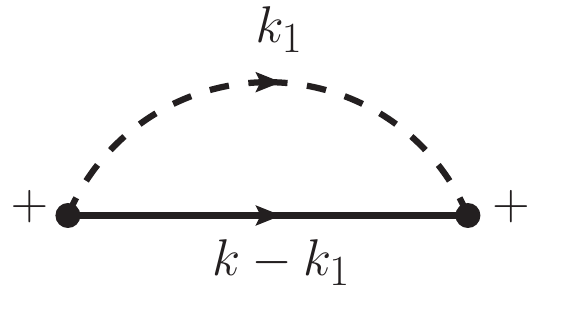}

\caption{The Feynman diagram for pion self-energy $\Sigma^{F}$ at LO. The
solid line represents the $\rho^{0}$ propagator and the dashed line
represents the pion propagator. \label{fig:Leading-order-Feynman}}
\end{figure}

The Feynman diagram for the pion self-energy $\Sigma^{F}$ at LO is
shown in Fig.(\ref{fig:Leading-order-Feynman}) which is given by
\begin{eqnarray}
-i\Sigma^{F}(k) & = & -g_{V}^{2}\int\frac{d^{4}k_{1}}{\left(2\pi\hbar\right)^{4}}S^{F}(k_{1})G_{\alpha\beta}^{F}(k-k_{1})\left(k+k_{1}\right)^{\alpha}\left(k+k_{1}\right)^{\beta}.\label{eq:pi_self-energy_F}
\end{eqnarray}
where the Feynman propagators in medium read
\begin{eqnarray}
S^{F}(k) & = & \frac{i}{k^{2}-m_{\pi}^{2}+i\epsilon}+2\pi\hbar\delta\left(k^{2}-m_{\pi}^{2}\right)\left[\theta(k^{0})f_{\pi^{+}}(\mathbf{k})+\theta(-k^{0})f_{\pi^{-}}(-\mathbf{k})\right],\label{eq:pi_Feynman_propogator}\\
G_{\alpha\beta}^{F}(p) & = & -\frac{i\left(g_{\alpha\beta}-p_{\alpha}p_{\beta}/m_{\rho}^{2}\right)}{p^{2}-m_{\rho}^{2}+i\epsilon}+\left(2\pi\hbar\right)\delta\left(p^{2}-m_{\rho}^{2}\right)\nonumber \\
 &  & \times\left[\theta(p^{0})\epsilon_{\alpha}\left(s_{1},{\bf p}\right)\epsilon_{\beta}^{*}\left(s_{2},{\bf p}\right)f_{s_{1}s_{2}}(\mathbf{p})+\theta(-p^{0})\epsilon_{\alpha}^{\ast}\left(s_{1},-{\bf p}\right)\epsilon_{\beta}\left(s_{2},-{\bf p}\right)f_{s_{2}s_{1}}(-\mathbf{p})\right],\label{eq:rho_Feynman_propogator}
\end{eqnarray}
which can be derived by substituting Eqs. (\ref{eq:vector field})
and (\ref{eq:scalar field}) into Eqs. (\ref{eq:G_vector}) and (\ref{eq:G_scalar}).

Substituting Eqs. (\ref{eq:pi_Feynman_propogator}), (\ref{eq:rho_Feynman_propogator})
into Eq. (\ref{eq:pi_self-energy_F}), we obtain the imaginary part
of the self-energy
\begin{eqnarray}
\Gamma(k) & \equiv & \mathrm{Im}\Sigma^{F}(k)=2g_{V}^{2}\theta(k^{0})\int\frac{d^{3}k_{1}}{(2\pi\hbar)^{3}2E_{k_{1}}^{\pi}}\int\frac{d^{3}p}{(2\pi\hbar)^{3}2E_{p}^{\rho}}\nonumber \\
 &  & \times(2\pi\hbar)^{4}\delta^{(4)}\left(k+k_{1}-p\right)f_{\pi^{-}}(\mathbf{k}_{1})\left[m_{\pi}^{2}-\frac{\left(k_{1}\cdot p\right)^{2}}{m_{\rho}^{2}}\right]\nonumber \\
 &  & +2g_{V}^{2}\theta(-k^{0})\int\frac{d^{3}k_{1}}{(2\pi\hbar)^{3}2E_{k_{1}}^{\pi}}\int\frac{d^{3}p}{(2\pi\hbar)^{3}2E_{p}^{\rho}}\nonumber \\
 &  & \times(2\pi\hbar)^{4}\delta^{(4)}\left(k-k_{1}+p\right)f_{\pi^{+}}(\mathbf{k}_{1})\left[m_{\pi}^{2}-\frac{\left(k_{1}\cdot p\right)^{2}}{m_{\rho}^{2}}\right],\label{eq:im-part}
\end{eqnarray}
where we have assumed that $k$ is near the mass-shell, since the
self-energy's correction to $k^{2}-m_{\pi}^{2}$ in Eq. (\ref{eq:Feynman-propogator-F})
is negligible if $k$ is far off-shell. Under such an assumption,
processes such as $\pi^{+}\rightarrow\pi^{+}\rho^{0}$ are forbidden,
so the self-energy can be simplified. With the imaginary part of the
self-energy in (\ref{eq:im-part}), the function $D_{(1)}(s,\lambda)$
in $C_{\mathrm{scat}}\left(\rho^{0}\pi^{\pm}\leftrightarrow\rho^{0}\pi^{\pm}\right)$
in Eq. (\ref{eq:pi-propagator}) becomes
\begin{align}
D_{\pi^{+}(1)}(s,\lambda)= & \frac{\left[k_{1}\cdot\epsilon\left(s,\mathbf{p}_{1}\right)\right]\left[k_{2}\cdot\epsilon^{*}\left(\lambda,{\bf p}\right)\right]}{\left(p+k_{2}\right)^{2}-m_{\pi}^{2}+i\Gamma(p+k_{2})}+\frac{\left[k_{2}\cdot\epsilon\left(s,\mathbf{p}_{1}\right)\right]\left[k_{1}\cdot\epsilon^{*}\left(\lambda,{\bf p}\right)\right]}{\left(p-k_{1}\right)^{2}-m_{\pi}^{2}+i\Gamma(-p+k_{1})},\nonumber \\
D_{\pi^{-}(1)}(s,\lambda)= & \frac{\left[k_{1}\cdot\epsilon\left(s,\mathbf{p}_{1}\right)\right]\left[k_{2}\cdot\epsilon^{*}\left(\lambda,{\bf p}\right)\right]}{\left(p+k_{2}\right)^{2}-m_{\pi}^{2}+i\Gamma(-p-k_{2})}+\frac{\left[k_{2}\cdot\epsilon\left(s,\mathbf{p}_{1}\right)\right]\left[k_{1}\cdot\epsilon^{*}\left(\lambda,{\bf p}\right)\right]}{\left(p-k_{1}\right)^{2}-m_{\pi}^{2}+i\Gamma(p-k_{1})}.\label{eq:pi-propagator-1}
\end{align}
which are different for $\rho^{0}\pi^{+}\leftrightarrow\rho^{0}\pi^{+}$
and $\rho^{0}\pi^{-}\leftrightarrow\rho^{0}\pi^{-}$ processes.


\section{Numerical results \label{sec:Numerical-result}}

\subsection{Initial condition without elliptic flow}

Since we are studying the spin alignment of $\rho^{0}$ in a pion
gas, we assume the pion density is much larger than the density of
$\rho^{0}$, $f_{\lambda_{1}\lambda_{2}}\ll f_{\pi^{\pm}}$, so the
influence of $\rho^{0}$ mesons on pions is negligible. We further
assume that $\pi^{\pm}$ are in global thermal equilibrium, so they
obey the Bose-Einstein distribution
\begin{equation}
f_{\pi^{\pm}}(x,\mathbf{p})=f_{\pi^{\pm}}(\mathbf{p})=\frac{1}{\exp\left[\beta\left(E_{p}\mp\mu_{\pi}\right)\right]-1},
\end{equation}
where $\beta=1/T$ is the inverse temperature, $\mu_{\pi}$ is the
chemical potential for $\pi^{+}$. Here we neglected the spatial dependence
of distributions. We choose $\mu_{\pi}$=0, and $T=156.5$ MeV corresponding
to the chemical freezeout temperature. Because $f_{\lambda_{1}\lambda_{2}}\ll f_{\pi^{\pm}}$
we can neglect the terms of order $f_{\lambda_{1}\lambda_{2}}^{2}$
relative to $f_{\lambda_{1}\lambda_{2}}$. Since the temperature is
much less than $m_{\rho}$, the contribution from the process $\rho^{0}\rho^{0}\leftrightarrow\pi^{+}\pi^{-}$
is negligible (two orders of magnitude smaller) relative to $C_{\mathrm{coal}/\mathrm{diss}}^{(0)}\left(\rho^{0}\leftrightarrow\pi^{+}\pi^{-}\right)$.

In summary, the collision terms that we take into account are $C_{\mathrm{coal}/\mathrm{diss}}^{(0)}\left(\rho^{0}\leftrightarrow\pi^{+}\pi^{-}\right)$
and $C_{\mathrm{scat}}\left(\rho^{0}\pi^{\pm}\leftrightarrow\rho^{0}\pi^{\pm}\right)$.
For $f_{\pi^{+}}=f_{\pi^{-}}$, we can simply have $C_{\mathrm{scat}}\left(\rho^{0}\pi^{+}\leftrightarrow\rho^{0}\pi^{+}\right)=C_{\mathrm{scat}}\left(\rho^{0}\pi^{-}\leftrightarrow\rho^{0}\pi^{-}\right)$.


Considering the spin Boltzmann equation (\ref{eq:MVSD-Boltzmann-equation})
is an integral-differential equation, we use Monte Carlo method to
solve it. We build a 50$\times$50$\times$50 lattice in momentum
space for $\rho^{0}$ with lattice cell size 100$\times$100$\times$100
MeV$^{3}$, so the range $p_{x}$, $p_{y}$ and $p_{z}$ is $\left[-2.5,2.5\right]$
GeV, which is big enough compared with the temperature. The value
of $\rho_{00}=f_{00}/\mathrm{Tr}(f)$ represents the spin alignment
of $\rho^{0}$ mesons.

\begin{figure}
\includegraphics[scale=0.5]{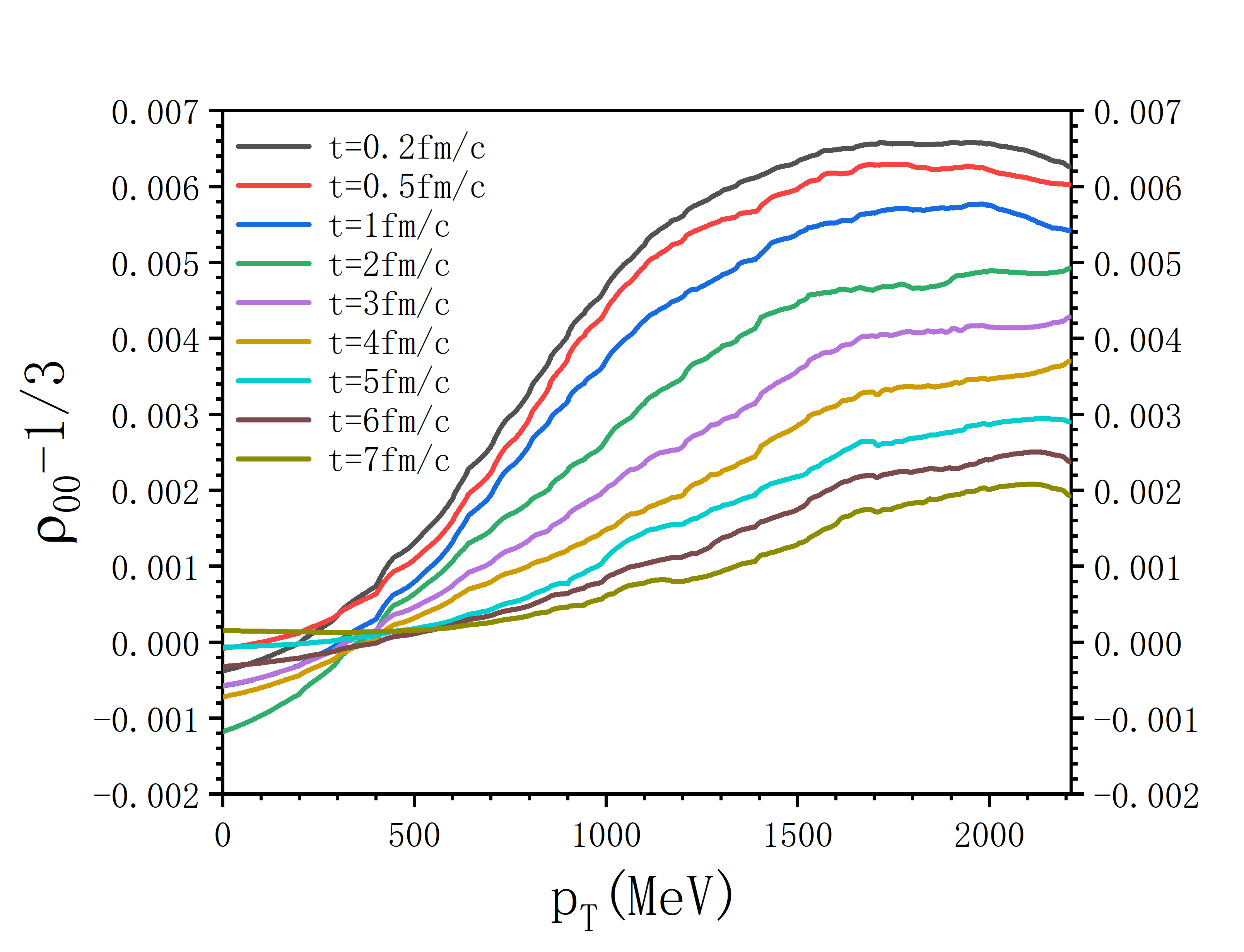}

\caption{The spin alignment as functions of $p_{T}$ at different time. The
initial distribution of rho mesons is set to $f_{\lambda_{1}\lambda_{2}}(t=0)=0$.
\label{fig:spin-alignment-without-initial}}
\end{figure}
\begin{figure}
\includegraphics[scale=0.5]{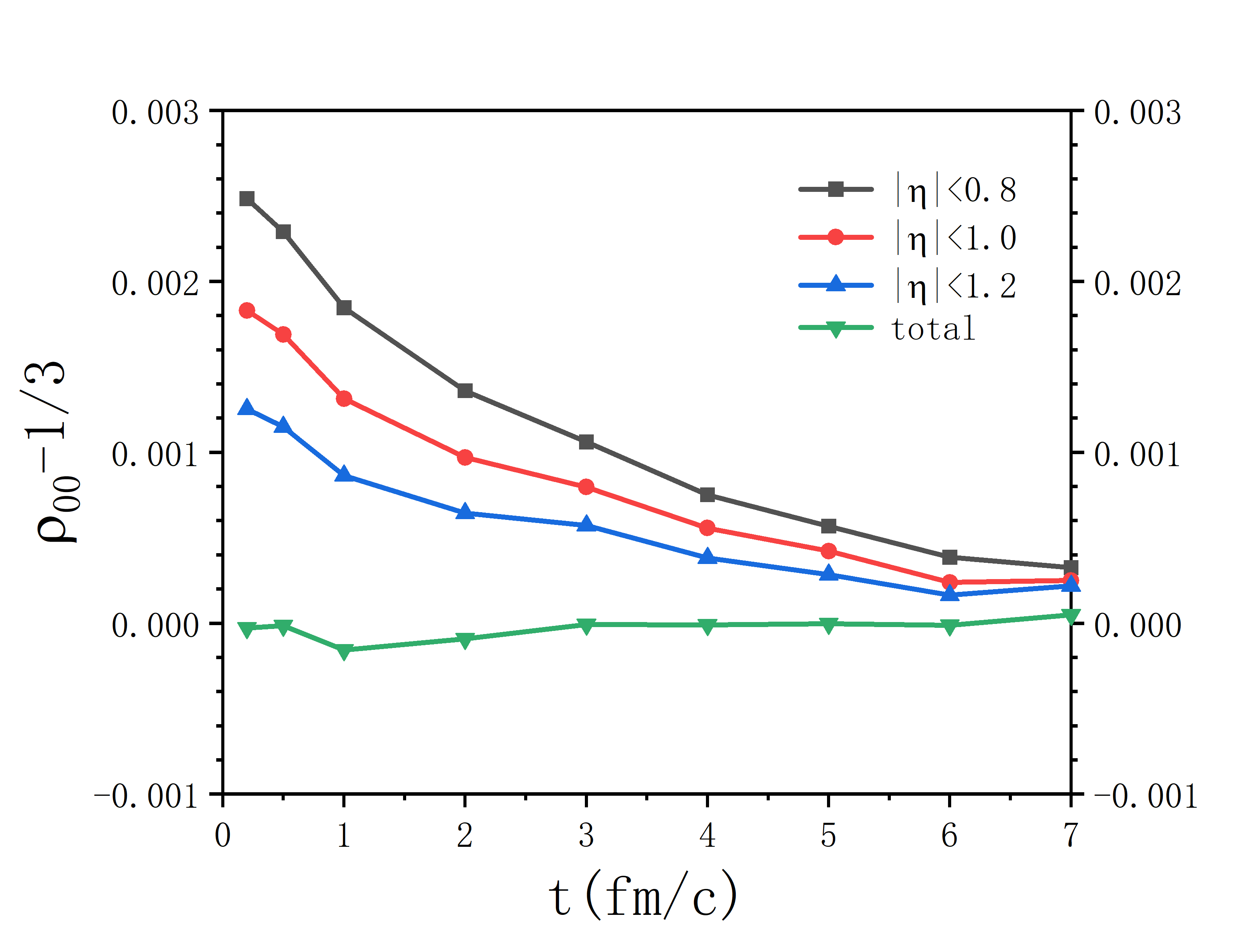}

\caption{The $p_{T}$-integrated spin alignment in different pseudorapidity
ranges for the initial distribution $f_{\lambda_{1}\lambda_{2}}(t=0)=0$.
\label{fig:The-spin-alignment-1}}

\end{figure}

In the first case, we consider the initial condition without neutral
rho mesons, i.e. $f_{\lambda_{1}\lambda_{2}}(t=0)=0$. The time step
for simulation is chosen to be $5\times10^{-6}$ MeV$^{-1}$$\approx10^{-3}$
fm/c. The spin alignments of rho mesons as functions of $p_{T}$ in
the pseudorapidity range $|\eta|<1$ at different time are shown in
Fig. (\ref{fig:spin-alignment-without-initial}). The spin alignments
($p_{T}$ integrated) in different pseudorapidity ranges are shown
in Fig. (\ref{fig:The-spin-alignment-1}). The precision of $\rho_{00}$
is about $10^{-3}$ in Monte Carlo method, so the results less than
$10^{-3}$ are not reliable. However, we can still see the time and
pseudorapidity dependence of the spin alignment from these results.


We notice that $\rho_{00}$ is slightly larger than 1/3 in the central
rapidity region of rho mesons though the pion distribution is isotropic.
It is because that we choose $+y$ to be the spin quantization direction,
which is different from $x$ and $z$. More specifically, the produced
rho mesons with momenta in $\pm y$ direction have $\rho_{00}>1/3$,
while those with momenta near the $xz$ plane have $\rho_{00}<1/3$.
The spin alignment in the whole momentum space must be zero because
of the isotropic pion distribution and angular momentum conservation,
as shown by the green line in Fig. (\ref{fig:The-spin-alignment-1}).
Therefore if we exclude rho mesons with momenta near $\pm z$ direction,
i.e. the forward and backward rapidity region, we have $\rho_{00}>1/3$.
The larger central pseudorapidity range we choose, the smaller spin
alignment we obtain. Since the scattering term contributes significantly
to a thermalization effect, we notice that the spin alignment decreases
rapidly with time.

\begin{figure}
\includegraphics[scale=0.5]{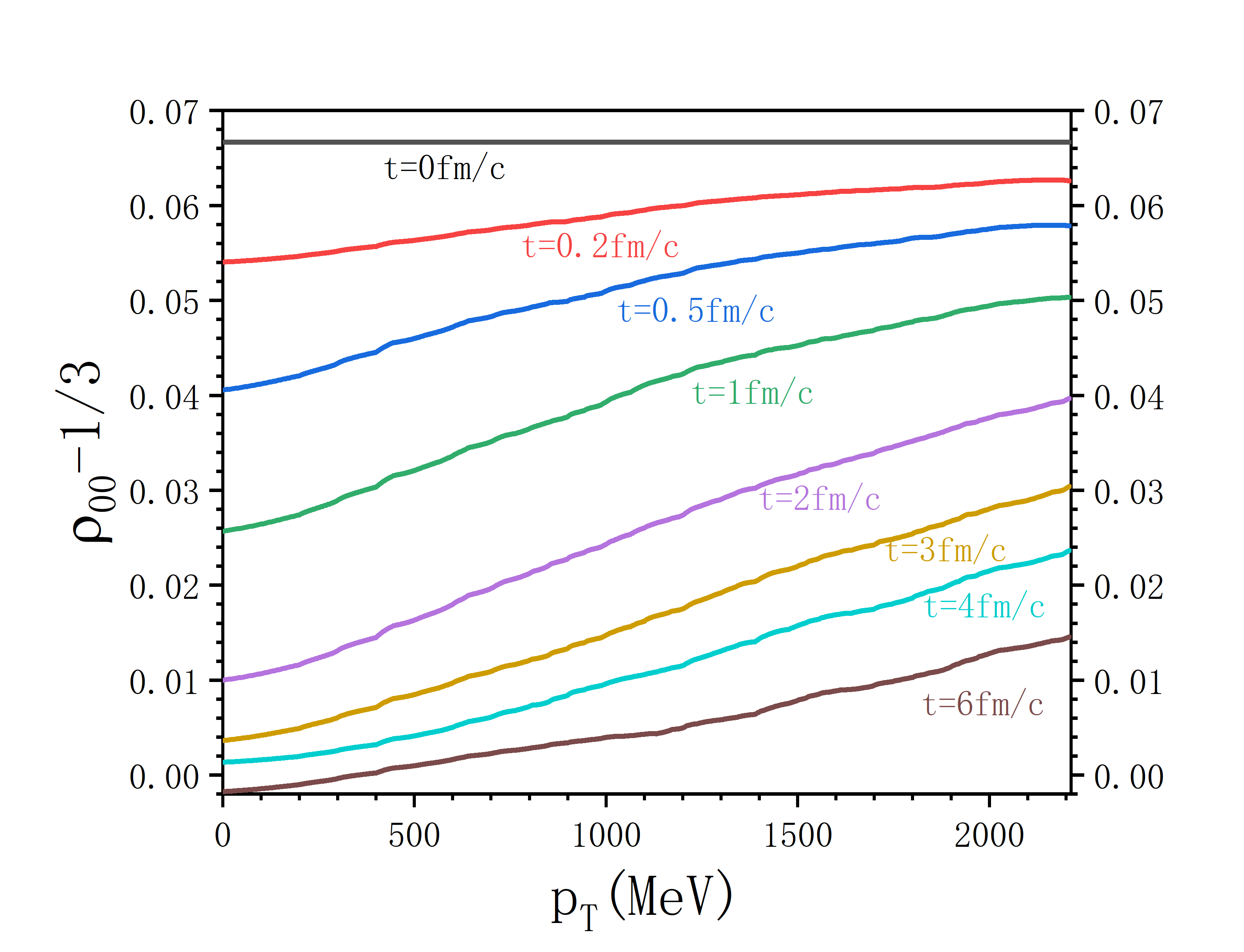}

\caption{The spin alignment as functions of $p_{T}$ in $|\eta|<1$ at different
time with the initial distribution (\ref{eq:NE-initial-condition})
that corresponds to $\rho_{00}=0.4>1/3$. \label{fig:spin-alignment}}
\end{figure}
\begin{figure}
\includegraphics[scale=0.5]{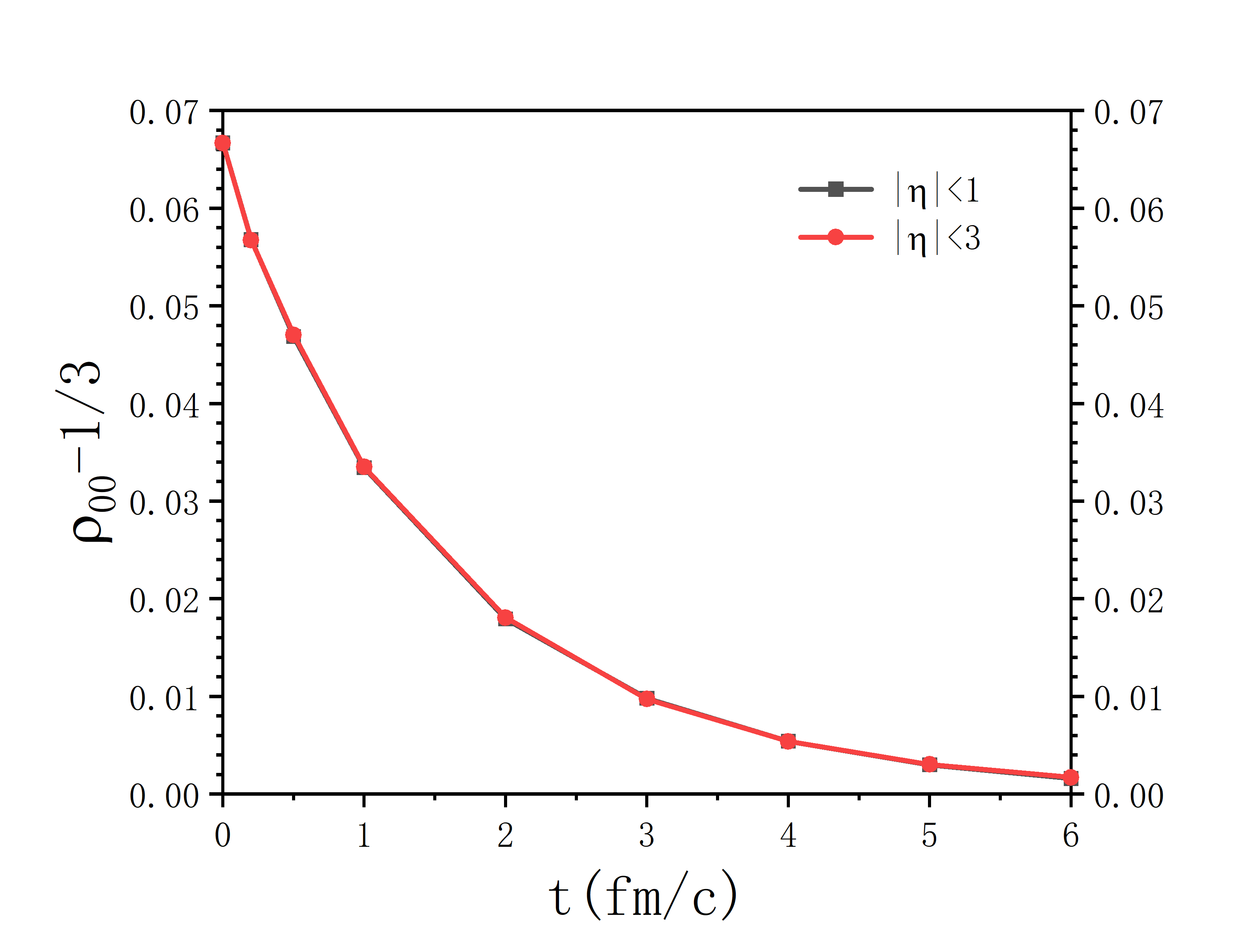}

\caption{The $p_{T}$-integrated spin alignment in different pseudorapidity
ranges with the initial distribution (\ref{eq:NE-initial-condition}).
\label{fig:spin-alignment-2}}
\end{figure}

In the second case, we consider a more realistic initial condition
by assuming an initial value of the spin alignment at the hadronization
time when the rho meson is formed by recombination of quarks. We set
the initial distribution of the rho meson as a thermal distribution
with the spin alignment $\rho_{00}=0.4$ (larger than 1/3), then the
matrix valued spin distribution is put into the form
\begin{eqnarray}
f_{\lambda_{1}\lambda_{2}} & = & \mathrm{diag}(0.9,1.2,0.9)\times f_{\mathrm{BE}},\label{eq:NE-initial-condition}
\end{eqnarray}
where $f_{\mathrm{BE}}$ is the Bose-Einstein distribution for the
rho meson with zero chemical potential. The time step for simulation
is chosen to be $5\times10^{-5}$ MeV$^{-1}\approx$0.01 fm/c. In
the pseudorapidity range $|\eta|<1$, the numerical results for the
spin alignment as functions of $p_{T}$ at different time are shown
in Fig. (\ref{fig:spin-alignment}). The results for the $p_{T}$-integrated
spin alignment in different pseudorapidity ranges are shown in Fig.
(\ref{fig:spin-alignment-2}). We can see that the spin alignment
is almost independent of the pseudorapidity range, because it is mostly
contributed from initial rho mesons with non-vanishing spin alignment
instead of from newly generated rho mesons. More importantly, we see
that $\rho_{00}-1/3$ decreases rapidly from the initial value 0.066
to 0.006 at $t=4$ fm/c, meaning that the initial value of the spin
alignment can be easily washed out by the interaction between rho
mesons and pions.


\begin{figure}
\includegraphics[scale=0.5]{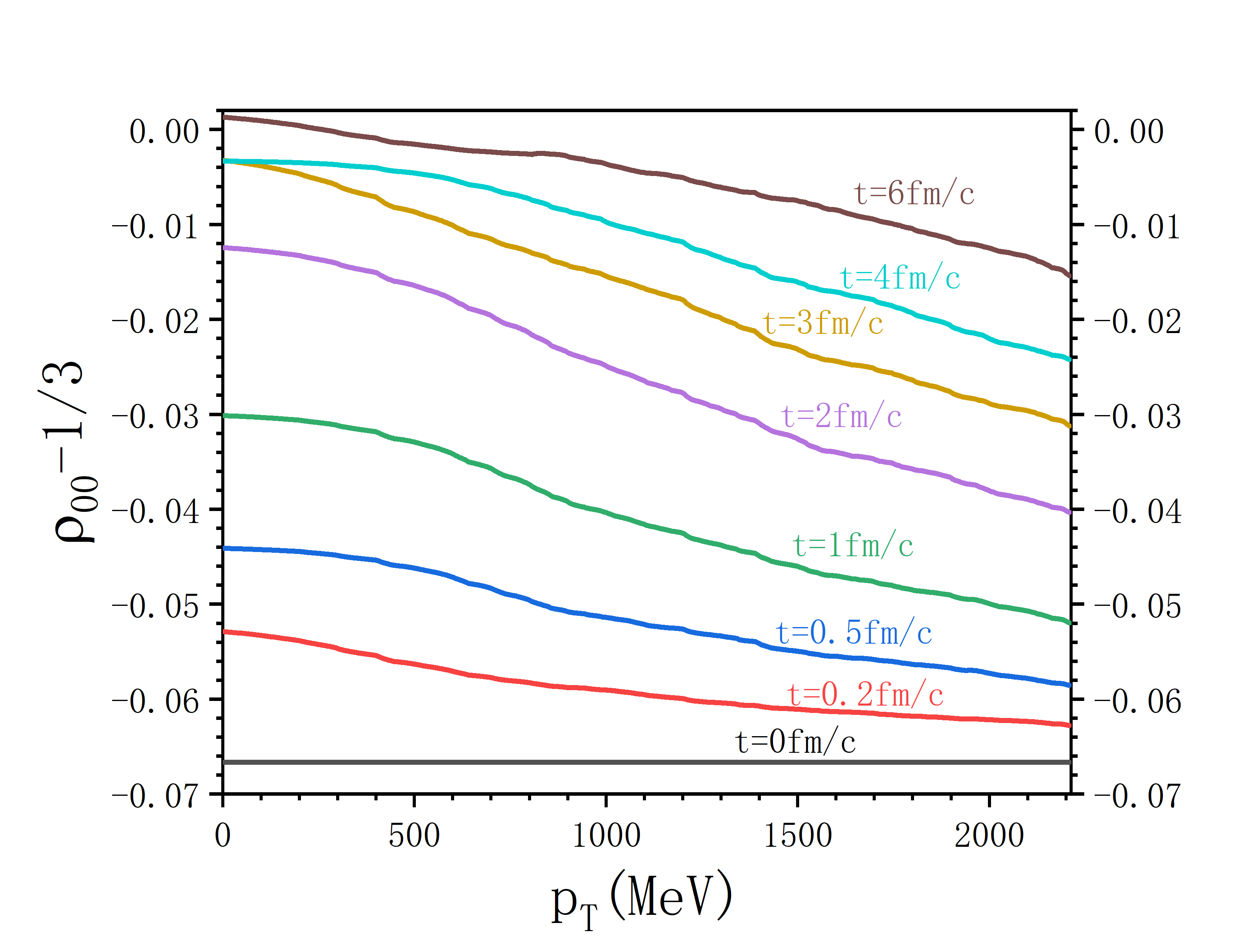}

\caption{The spin alignment as functions of $p_{T}$ at different time with
the initial distribution (\ref{eq:Initial_distribution_less}) that
corresponds to $\rho_{00}=0.27<1/3$. \label{fig:The-spin-alignment-less}}
\end{figure}

\begin{figure}
\includegraphics[scale=0.5]{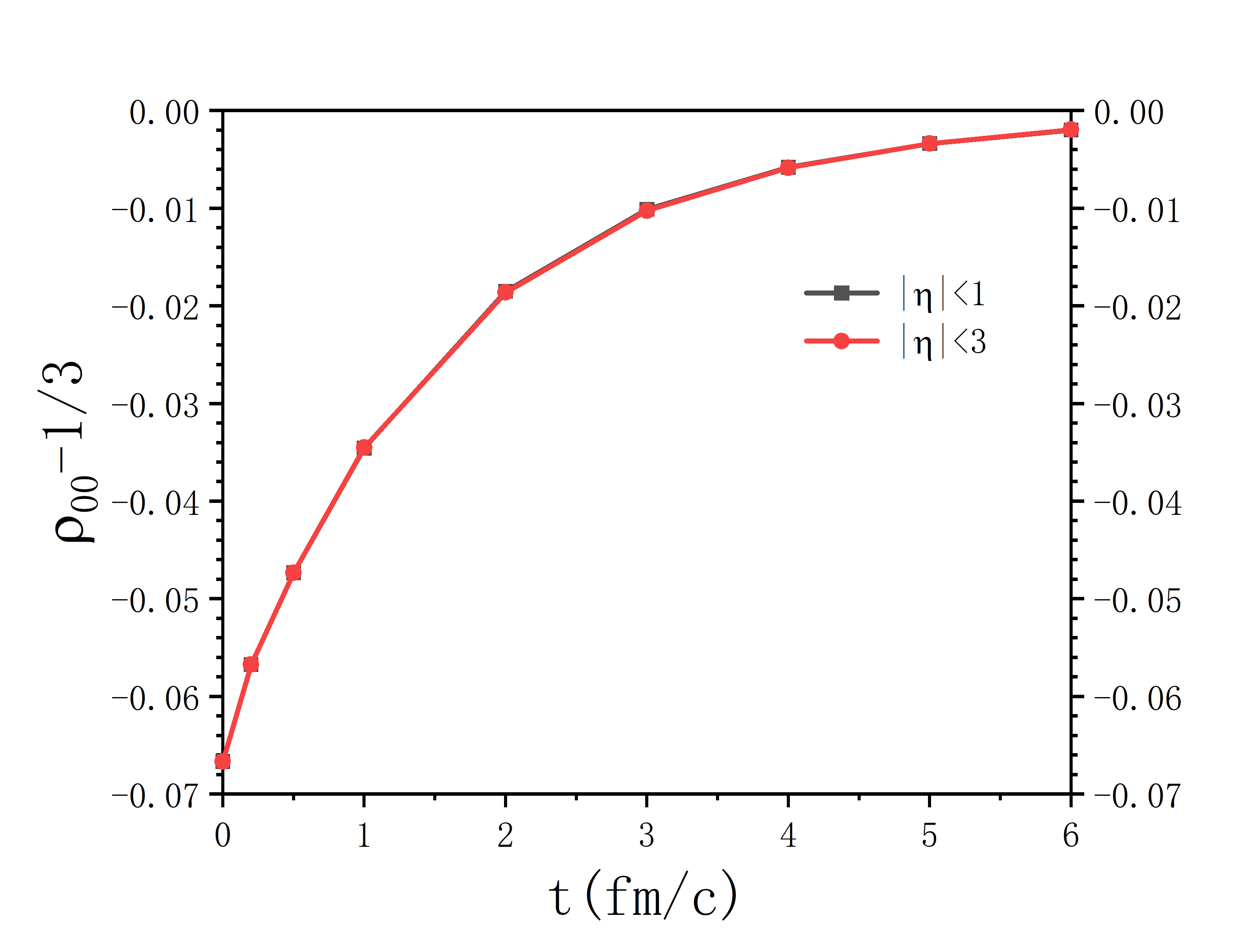}

\caption{The $p_{T}$-integrated spin alignment in different pseudorapidity
ranges with the initial distribution (\ref{eq:Initial_distribution_less}).\label{fig:The--integrated-spin-less-2}}

\end{figure}

We can also consider $\rho_{00}=0.27$ (less than 1/3) at the initial
time. Then the matrix valued spin distribution is set to
\begin{eqnarray}
f_{\lambda_{1}\lambda_{2}} & = & \mathrm{diag}(1.1,0.8,1.1)\times f_{\mathrm{BE}}.\label{eq:Initial_distribution_less}
\end{eqnarray}
The results are shown in Figs. (\ref{fig:The-spin-alignment-less})
and (\ref{fig:The--integrated-spin-less-2}). We see that the spin
alignment relaxes to 1/3 rapidly.

\subsection{Initial condition with elliptic flow}

In order to see the $v_{2}$ influence on the spin alignment of $\rho^{0}$,
we use the blast wave model \citep{Bondorf:1978kz,Siemens:1978pb,Schnedermann:1993ws,Retiere:2003kf}
to describe the space-time evolution of the fireball in heavy-ion
collisions. The idea is as follows. We assume Eq. (\ref{eq:boltzmann-compact})
describes the time evolution of $f_{\lambda_{1}\lambda_{2}}(x,\mathbf{p})$
in the fluid element's comoving frame located at $x$. The fluid four-velocity
$u^{\mu}(x)$ is described by the blast wave model for the boost invariant
expansion of the fireball along $z$ direction. The emission function
of the blast wave model has the form \citep{Retiere:2003kf}
\begin{eqnarray}
S(r,\phi_{s},p) & = & \theta(R-r)F(u,p),\label{eq:emission function}
\end{eqnarray}
where $R$ is the fireball's radius, $r$ and $\phi_{s}$ are the
radial position and the azimutal angle inside the fireball, $p$ is
the particle's momentum, $F(u,p)$ is some kind of the momentum distribution
function depending on the fluid velocity that can be parameterized
as
\begin{equation}
u^{\mu}(r,\phi_{s})=\left(\cosh\rho(r,\phi_{s}),\sinh\rho(r,\phi_{s})\cos\phi_{s},\sinh\rho(r,\phi_{s})\sin\phi_{s},0\right),
\end{equation}
where the radial flow rapidity $\rho$ is given by
\begin{eqnarray}
\rho(r,\phi_{s}) & = & \frac{r}{R}\left[\rho_{0}+\rho_{2}\cos(2\phi_{s})\right].
\end{eqnarray}
Here $\rho_{0}$ and $\rho_{2}$ are two parameters, and $\rho_{2}$
gives the elliptic flow. Note that without loss of generality we have
set space-time rapidity zero in $u^{\mu}(r,\phi_{s})$ corresponding
to $z=0$.

\begin{figure}
\includegraphics[scale=0.5]{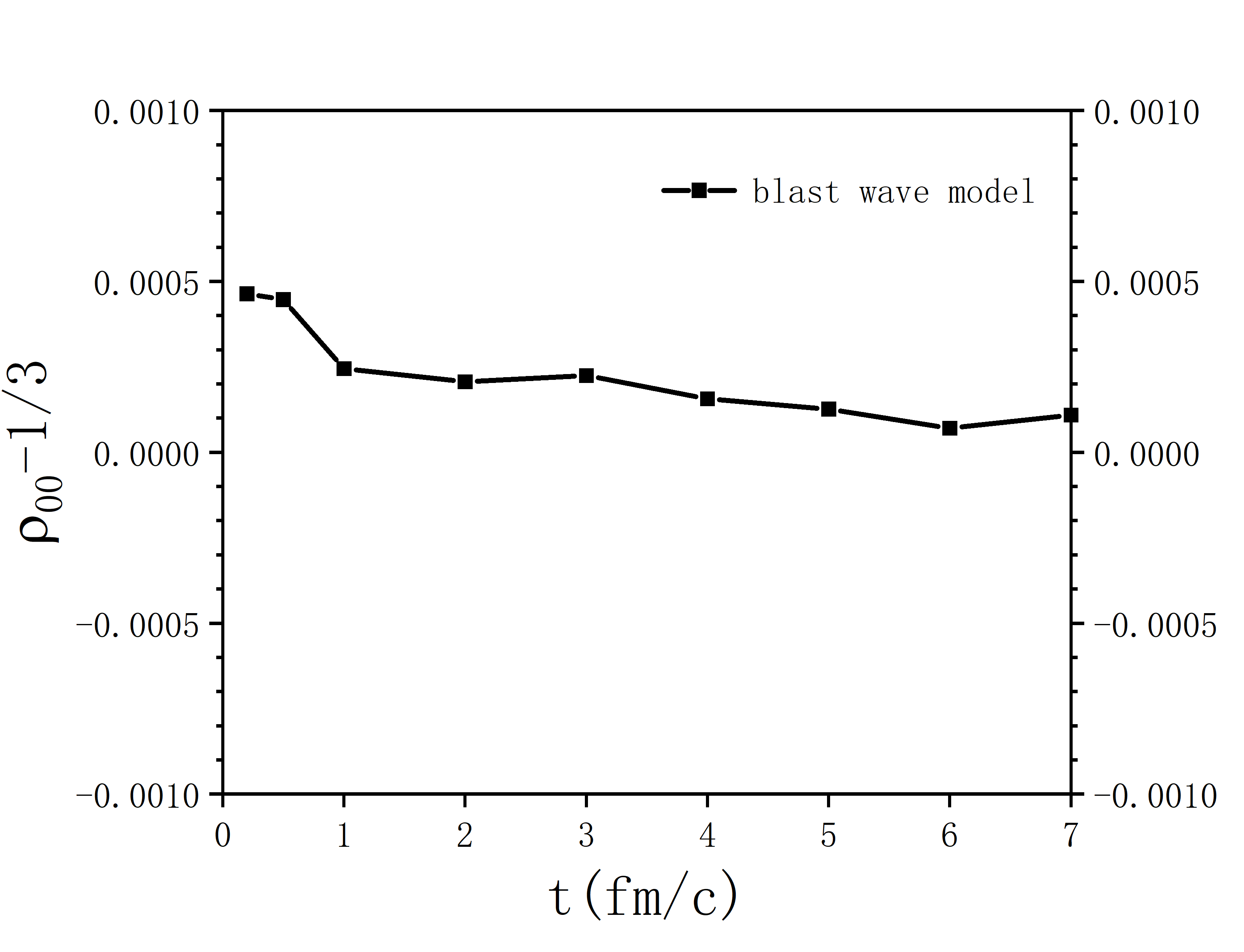}

\caption{The spin alignment of the neutral rho meson at $z=0$ for $|\eta|<1$
in the blast wave model with the elliptic flow. \label{fig:spin-blast-wave}}
\end{figure}

The parameters are chosen as $R=13$ fm, $\rho_{0}=0.89$, $\rho_{2}=0.06$
\citep{Retiere:2003kf}. We assume $f_{\lambda_{1}\lambda_{2}}=0$
at the initial time. Then with Eq. (\ref{eq:emission function}) and
these parameters we can calculate the spin alignment at $z=0$ as
follows
\begin{eqnarray}
\rho_{00} & = & \frac{\int_{|\eta|<1}d^{3}p\int_{0}^{R}rdrd\phi_{s}f_{00}(u,p)}{\int_{|\eta|<1}d^{3}p\int_{0}^{R}rdrd\phi_{s}\mathrm{tr}f(u,p)},\label{eq:blast-wave-rho00}
\end{eqnarray}
where we set $F(u,p)$ to $f_{\lambda_{1}\lambda_{2}}(u,p)$. It is
obvious that $\rho_{00}$ in Eq. (\ref{eq:blast-wave-rho00}) encodes
the effect of the elliptic flow. The results for $\rho_{00}$ are
shown in Fig. \ref{fig:spin-blast-wave} indicating that its deviation
from 1/3 is positive but in the order of $10^{-4}$.


\section{Conclusions and discussions}

\label{sec:Conclusions-and-discussions}Using two-point Green's functions
and Kadanoff-Baym equation in the closed-time path formalism for vector
mesons developed in the previous work \citep{Sheng:2022ffb}, we derived
spin kinetic or Boltzmann equations for neutral rho mesons in a pion
gas. The $\rho\pi\pi$ coupling is described by the chiral effective
theory. The collision terms in the pion gas at the leading and next-to-leading
order are obtained.\textcolor{red}{{} }We simulated the evolution of
the matrix valued spin distribution (spin density matrix) of neutral
rho mesons by the Monte Carlo method. In the simulation, we have assumed
the Bose-Einstein distribution for pions with $T=156.5$ MeV and vanishing
chemical potential. The numerical results show that the interaction
of pions and neutral rho mesons creates very small spin alignment
for rho mesons in the central rapidity region if there is no rho meson
in the system at the initial time. But there is no spin alignment
in the full rapidity range since pions' momenta are isotropic. Such
a small spin alignment in the central rapidity region will decay rapidly
toward zero in later time. If there are rho mesons with a sizable
spin alignment at the initial time the spin alignment will also decrease
rapidly. We also considered the effect on $\rho_{00}$ from the elliptic
flow of pions in the blast wave model. With vanishing spin alignment
at the initial time, the deviation of $\rho_{00}$ from 1/3 is positive
but very small.

The work can be improved or extended by loosening some approximations
or restrictions. For example, we can consider fluctuations in the
temperature and the distribution of pions in collision terms, or we
can consider other vector mesons in a hadrons gas. These can be done
in the future.

\begin{acknowledgments}
We thank A,-H. Tang for suggesting this topic for us and for insightful
discussion. We thank J.-H. Gao, X.-G. Huang, S. Lin, E. Speranza,
D. Wagner, D.-L. Yang for helpful discussion. This work is supported
in part by the Strategic Priority Research Program of the Chinese
Academy of Sciences (CAS) under Grant No. XDB34030102, and by the
National Natural Science Foundation of China (NSFC) under Grant No.
12135011 and 12075235.
\end{acknowledgments}

\bibliographystyle{h-physrev}
\phantomsection\addcontentsline{toc}{section}{\refname}\bibliography{citation-1}

\end{document}